\documentclass[aps, 10pt, prd,
               notitlepage, twocolumn, superscriptaddress, longbibliography,
               nofootinbib, floatfix]{revtex4-2}

\usepackage{amsmath}
\usepackage{amssymb}
\usepackage{amsfonts}
\usepackage[utf8]{inputenc}
\usepackage[T1]{fontenc}
\usepackage{mathrsfs}
\usepackage{anyfontsize}
\usepackage{stmaryrd}
\usepackage{dsfont}
\usepackage{microtype}
\usepackage{tensor}
\usepackage{bm}

\usepackage{graphicx}
\usepackage{epsfig}
\usepackage{epstopdf}

\usepackage[usenames,dvipsnames]{xcolor}
\usepackage[linktocpage,breaklinks]{hyperref}

\hypersetup{colorlinks=true,
            citecolor=NavyBlue,
            linkcolor=NavyBlue,
            urlcolor=NavyBlue}

\usepackage[normalem]{ulem}

\newcommand{\dd}{{\rm d}}
\newcommand{\pd}{\partial}
\newcommand{\vp}{\varphi}
\newcommand{\sG}{\mathcal{G}}
\newcommand{\sS}{\mathscr{S}}

\newcommand{\sA}{\mathscr{A}}

\newcommand{\cn}[1]{{\footnotesize \textcircled{\scriptsize #1}}}

\newcommand{\mx}{{\mbox{\tiny max}}}

\newcommand{\vpbg}{\bar{\varphi}}

\newcommand{\lstar}{\ell_{\ast}}
\newcommand{\lsqr}{\ell^{2}}

\newcommand{\aei}{Max Planck Institute for Gravitational Physics (Albert Einstein Institute),
\\ Am M\"uhlenberg 1, 14476 Potsdam, Germany}
\newcommand{\jhu}{Department of Physics and Astronomy, Johns Hopkins University,
\\ 3400 North Charles Street, Baltimore, Maryland 21218, USA}
\newcommand{\uiuc}{Illinois  Center  for  Advanced  Studies  of  the  Universe, Department of Physics,
\\ University of Illinois at Urbana-Champaign, Urbana, Illinois 61801, USA}

\begin{document}
\title{Black hole sensitivities in Einstein-scalar-Gauss-Bonnet gravity}

\date{\today}

\begin{abstract}
The post-Newtonian dynamics of black hole binaries in Einstein-scalar-Gauss-Bonnet
theories of gravity depends on the so-called ``sensitivities'', quantities
which characterize a black hole's adiabatic response to the time-dependent
scalar field environment sourced by its companion.
In this work, we calculate numerically the sensitivities of nonrotating black
holes, including spontaneously scalarized ones, in three classes of Einstein-scalar-Gauss-Bonnet gravity: the shift-symmetric, dilatonic and Gaussian theories.
When possible, we compare our results against perturbative analytical results, finding
excellent agreement.
Unlike their general relativistic counterparts, black holes in
Einstein-scalar-Gauss-Bonnet gravity only exist in a restricted parameter space controlled by the theory's coupling constant.
A preliminary study of the role played by the sensitivities in black hole
binaries suggests that, in principle, black holes can be driven outside of
their domain of existence during the inspiral, for binary parameters which we
determine.
\end{abstract}

\author{F\'elix-Louis Juli\'e}
\affiliation{\aei}
\affiliation{\jhu}

\author{Hector O. Silva}
\affiliation{\aei}
\affiliation{\uiuc}

\author{Emanuele Berti}
\affiliation{\jhu}

\author{Nicol\'as Yunes}
\affiliation{\uiuc}

\maketitle

\section{Introduction}

The detection of gravitational waves from compact binary coalescences by the
LIGO-Virgo collaboration~\cite{LIGOScientific:2018mvr,LIGOScientific:2020ibl,LIGOScientific:2021djp}
started a new era in experimental gravitational
physics where, for the first time, we can test the predictions of general
relativity (and modifications thereof) in highly dynamical, nonlinear
environments~\cite{TheLIGOScientific:2016src,Yunes:2016jcc,LIGOScientific:2019fpa,LIGOScientific:2020tif,LIGOScientific:2021sio,Ghosh:2021mrv,Nair:2019iur,Perkins:2021mhb,Lyu:2022gdr}.
A prerequisite to perform such tests is a description of the orbital dynamics
and the associated gravitational wave emission of inspiralling compact objects
(i.e., neutron stars and black holes) in relativistic gravity theories~\cite{Berti:2015itd,Berti:2018cxi,Berti:2018vdi}.

A well-motivated class of modifications to general relativity introduces a
dynamical scalar field that couples nonminimally to the Gauss-Bonnet density.
These Einstein-scalar-Gauss-Bonnet (ESGB) theories arise in the low-energy limit of
heterotic string theory~\cite{Metsaev:1987zx}, and also from the
dimensional reduction of higher-dimensional Lovelock
theories~\cite{Charmousis:2014mia}.
They are a subclass of Horndeski
gravity~\cite{Kobayashi:2011nu,Kobayashi:2019hrl} and also arise from an
effective field theory perspective~\cite{Yagi:2015oca,Cano:2019ore}.
Due to the coupling between the scalar field and the Gauss-Bonnet
density, black holes in these theories can violate no-hair
theorems~\cite{Mignemi:1992nt,Kanti:1995vq,Guo:2008hf,Pani:2009wy,Yunes:2011we,Pani:2011gy,Sotiriou:2013qea,Sotiriou:2014pfa,Maselli:2015yva,Kleihaus:2015aje,Antoniou:2017acq,Prabhu:2018aun,Saravani:2019xwx,Sullivan:2019vyi,Sullivan:2020zpf,Delgado:2020rev}
and exhibit spontaneous
scalarization~\cite{Doneva:2017bvd,Silva:2017uqg,Dima:2020yac,Herdeiro:2020wei,Berti:2020kgk}.
As a consequence, black holes are endowed with a monopole scalar charge, which
can source dipolar scalar radiation in binary black hole
systems~\cite{Yagi:2011xp,Stein:2013wza}.
This makes black hole binaries ideal systems to constrain (or to look for evidence
in favor of) these theories with current~\cite{Nair:2019iur,Perkins:2021mhb,Lyu:2022gdr}
and future gravitational-wave observatories~\cite{Perkins:2020tra,Maselli:2020zgv}.

\begin{figure}[t]
\includegraphics[width=.9\columnwidth]{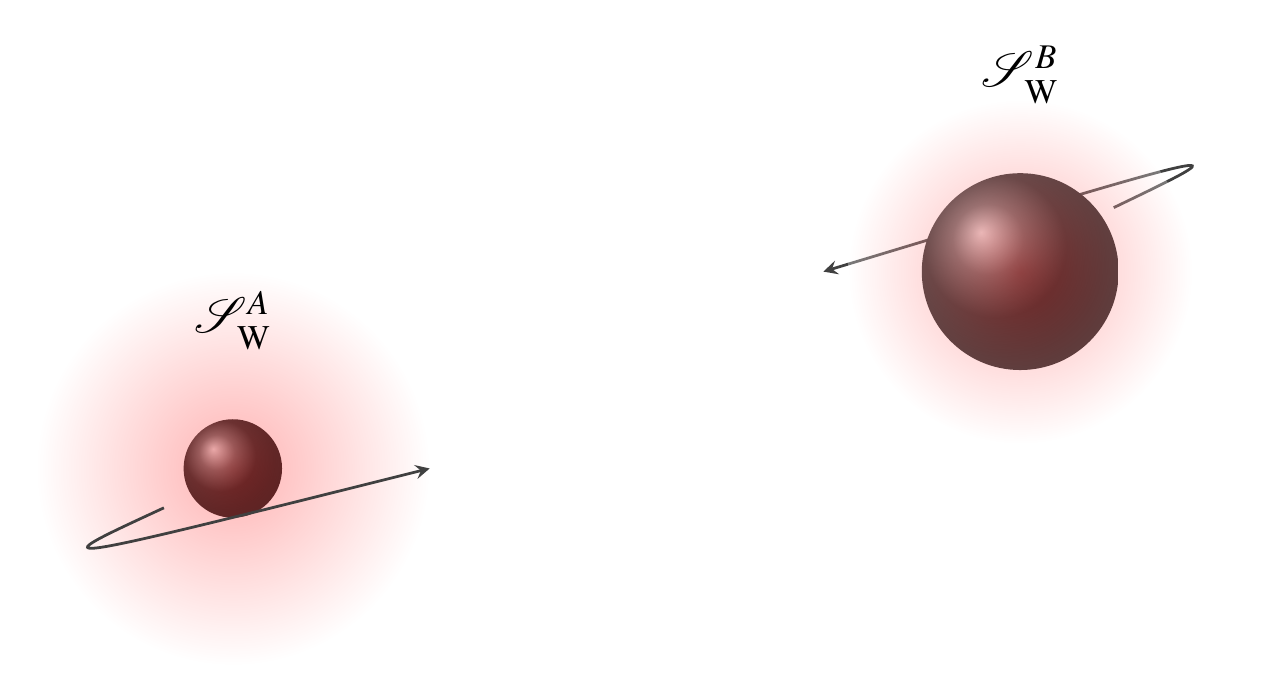}
\caption{Illustration of the problem. Two black holes with
Arnowitt-Misner-Deser (ADM) masses $M_{A,B}$ and with scalar charges $Q_{A,B}$ are in a binary system.
The scalar field of each black hole affects its companion, altering its mass and scalar charge.
In the PN regime, and when
finite-size corrections (e.g.~tidal and out-of-equilibrium effects) can be neglected, these changes take place adiabatically, keeping the
Wald entropy $\mathscr{S}_{\rm W}^{A,B}$ of each black hole constant.
The change in the mass due to a slowly-varying scalar field environment
at constant Wald entropy is the sensitivity, which we calculate here.}
\label{fig:bbh_diagram}
\end{figure}

With this motivation, considerable effort has been placed in developing tools
to model the dynamics of black hole
binaries in ESGB gravity, including the prediction of gravitational waveforms, using both post-Newtonian (PN)~\cite{Yagi:2011xp,Julie:2019sab,Shiralilou:2020gah,Shiralilou:2021mfl,Bernard:2022noq}
and numerical relativity~\cite{Witek:2018dmd,Ripley:2019hxt,Ripley:2019irj,Ripley:2019aqj,Okounkova:2020rqw,East:2020hgw,Silva:2020omi,East:2021bqk,Kuan:2021lol} approaches, the latter accompanied by works studying the
Cauchy problem in this theory~\cite{Papallo:2017qvl,Kovacs:2020ywu,Kovacs:2020pns,Julie:2020vov,Witek:2020uzz,Kovacs:2021lgk}.
The ``skeletonization'' of an analytical black hole solution in
this theory~\cite{Julie:2019sab} can be used to show that the two-body
Lagrangian describing the dynamics of black hole binaries at first PN order
requires the knowledge of the so-called ``sensitivities'', quantities which
characterize the adiabatic changes
(more precisely, at constant Wald entropy)
in the black holes' mass and scalar charge induced by the slowly
varying external scalar field sourced by their companion, as illustrated in Fig.~\ref{fig:bbh_diagram}.
These sensitivities are the black hole counterparts of a similar concept
introduced for neutron stars in scalar-tensor
theories in~\cite{Eardley:1975ApJ,Will:1989sk,Damour:1992we,Damour:1995kt,Damour:1996ke}.
They also arise in the two-body problem in
Einstein-Maxwell-scalar~\cite{Julie:2017rpw,Khalil:2018aaj,Julie:2018lfp} and in
Lorentz-violating theories~\cite{Yagi:2013ava,Gupta:2021vdj}.

Here we develop a method to compute the sensitivities of nonrotating black
holes using a full numerical approach. This requires, as a preliminary step, that we
obtain black hole solutions that generalize those in the literature: we obtain families of constant Wald entropy black holes with
nonvanishing asymptotic scalar fields.
The sensitivities were also calculated using the
analytical, but perturbative, approach of Ref.~\cite{Julie:2019sab}.
We show that the analytical Pad\'e-approximants
obtained there (and extended to higher orders here)
show remarkable agreement with numerical calculations.
We also calculate, for the first time, the sensitivity of nonrotating,
spontaneously scalarized black holes.
With these results at hand, we speculate that black holes with real and regular
scalar hair can cease to exist in binaries in ESGB theories, and
we discuss the implications of this possibility.
The method developed here to calculate the black hole sensitivity should
be applicable in other gravitational theories as well.

The paper is organized as follows.
In Sec.~\ref{sec:sgb_review} we briefly review ESGB gravity and
how black hole solutions are obtained numerically in this theory.
In Sec.~\ref{sec:bh_sens} we develop a strategy to calculate the sensitivities numerically,
and present results for selected classes of ESGB theories.
In Sec.~\ref{sec:application} we apply these results to study
the evolution of black holes in binaries in this theory.
Finally, in Sec.~\ref{sec:conclusions} we summarize our main findings and
discuss possible avenues for future work.
We use geometrical units ($G=c=1$) throughout this work.

\section{Einstein-scalar-Gauss-Bonnet gravity}
\label{sec:sgb_review}

\subsection{Action and field equations}

The theory we consider is described by the action
\begin{equation}
S = \frac{1}{16 \pi} \int \dd^{4}x\, \sqrt{-g}
\left[
R - 2 (\pd \vp)^2 + \lsqr f(\vp) \sG
\right]\,,
\label{eq:action}
\end{equation}
where we use the same notation as in Ref.~\cite{Julie:2019sab}: $R$ is the Ricci scalar, $g = {\rm det} \, g_{\mu\nu}$ is the metric
determinant, and $\vp$ a scalar field with kinetic term
$(\pd \vp)^2 = g^{\mu\nu} \pd_{\mu}\vp \pd_{\nu}\vp$
which couples to the Gauss-Bonnet invariant
\begin{align}
\sG = R^{\mu\nu\rho\sigma}R_{\mu\nu\rho\sigma} - 4R^{\mu\nu}R_{\mu\nu} + R^2
=R^{\mu\nu\rho\sigma}P_{\mu\nu\rho\sigma}\,,
\label{eq:gb}
\end{align}
where $R^{\mu}{}_{\nu\rho\sigma}$ and $R_{\mu\nu}$ are the Riemann and Ricci
tensors, respectively, and
\begin{align}
\!\! P^{\mu\nu}{}_{\rho\sigma} =
R^{\mu\nu}{}_{\rho\sigma}
- 2\delta^{\mu}{}_{[\rho} R^{\nu}{}_{\sigma]}
+ 2\delta^{\nu}{}_{[\rho} R^{\mu}{}_{\sigma]}
+ \delta^{\mu}{}_{[\rho} \delta^{\nu}{}_{\sigma]} R \,,
\nonumber\\
\label{eq:ptensor}
\end{align}
with brackets denoting antisymmetrization, as in
$\delta^{\mu}{}_{[\rho} \delta^{\nu}{}_{\sigma]} = (1/2)(\delta^{\mu}{}_{\rho} \delta^{\nu}{}_{\sigma}
- \delta^{\mu}{}_{\sigma} \delta^{\nu}{}_{\rho})$.
The tensor $ P_{\mu\nu\rho\sigma}$ has the symmetries of the Riemann tensor and
is divergence-free due to the Bianchi identities (see
e.g. Refs.~\cite{Deruelle:2003ps,Davis:2002gn,Julie:2020vov}).
The integral of the Gauss-Bonnet scalar over a four-dimensional spacetime $\int \dd^4x\sqrt{-g} \, \sG$ is a boundary term \cite{Myers:1987yn}.
The function $f(\vp)$ defines the theory, and the Gauss-Bonnet coupling
strength is set by the constant $\ell$, with dimensions of length.

The field equations of the theory, obtained by varying the
action~\eqref{eq:action} with respect to $g_{\mu\nu}$ and $\vp$, are:
\begin{subequations}
\begin{align}
\!\! R_{\mu\nu} &= 2 \pd_{\mu}\vp \pd_{\nu}\vp
- 4 \lsqr\! \left(P_{\mu\alpha\nu\beta} - \tfrac{1}{2}g_{\mu\nu} P_{\alpha\beta} \right)
\nabla^{\alpha} \nabla^{\beta} f\,,
\label{eq:eom_metric}\\
\Box \vp &= - (\lsqr\!/4) f_{,\vp}(\vp) \, \sG\,,
\label{eq:eom_scalar}
\end{align}\label{eq:eom}%
\end{subequations}
where $P_{\mu\nu} = P\indices{^\alpha_{\mu\alpha\nu}}$,
$\nabla_\mu$ is the metric-compatible covariant derivative associated to
$g_{\mu\nu}$, and $\Box=\nabla^\mu\nabla_\mu$.
We also use $(\cdot)_{,\vp} = \dd (\cdot) / \dd \vp$ to
indicate derivatives with respect to the scalar field $\vp$.

\subsection{Nonrotating black holes}
\label{subsec:nonrotating_bhs}

We are interested in obtaining static, spherically symmetric black hole
solutions.
We consider a line element of the form~\cite{Delgado:2020rev}:
\begin{align}
\dd s^2 &= -N(r)\sigma^2(r) \dd t^2 + N(r)^{-1} \dd r^2
\nonumber \\
&\quad + r^2 (\sin^2\theta \, \dd\theta^2 + \dd \phi^2)\,,
\label{eq:line_element}
\end{align}
in Schwarzschild-Droste coordinates $x^{\mu} = \{t,r,\theta,\phi \}$ and a scalar field $\vp(r)$.
We define
\begin{equation}
N(r) = 1 - 2 m(r)/r\,,
\label{eq:def_n}
\end{equation}
where $m(r)$ is the Misner-Sharp mass~\cite{Misner:1964je} such that
$m(r) \to M$ as $r \to \infty$, and $M$ is the ADM mass of the spacetime.
The Schwarzschild solution is recovered by setting
$m(r) = M$ and $\sigma =1$.
The Gauss-Bonnet invariant $\sG$ for this line element is
\begin{align}
\sG &= \frac{4}{r^2} \left[ {N'}^2
+ \frac{\sigma'}{\sigma} N'(5N-3)
\right.
\nonumber \\
&\quad\left.
+ N(N-1) \left( \frac{N''}{N} + \frac{2\sigma''}{\sigma} \right) \right] \,,
\label{eq:gb_for_line_element}
\end{align}
from which we can recover the familiar result $\sG=48 M^2 / r^6$ in the Schwarzschild limit.

For convenience, we introduce the dimensionless quantities
\begin{align}
r_{\ast} = r / r_{H}\,, \quad \textrm{and} \quad \lstar = \ell / r_{H}\,,
\end{align}
where for now $r_{H}$ is an arbitrary length (in this paper, an asterisk subscript will always denote a quantity that has been made dimensionless by dividing by $r_H$).
We can then use Eq.~\eqref{eq:line_element} in the field
equations~\eqref{eq:eom_metric}-\eqref{eq:eom_scalar} to obtain a system of
differential equations for $N'$, $\sigma'$, and $\varphi''$, where
$(\cdot)' = \dd (\cdot) / \dd r_{\ast}$.
The steps are as follows.
First, the $(tt)$- and $(rr)$-components of Eq.~\eqref{eq:eom_metric} provide
a system of two equations, which we can solve for $m'$ and $\sigma'$ in terms
only of $N$ and first and second derivatives of $\varphi$.  These are the first
two equations we need.
Second, from the $(\theta\theta)$-component of Eq.~\eqref{eq:eom_metric}, we
solve for $m''$.
Finally, we can use the equations obtained in the previous step to eliminate $m'$,
$m''$ and $\sigma'$ from Eq.~\eqref{eq:eom_scalar}. This is the third
equation we need.

Explicitly, the system of equations we work with is:
\begin{widetext}
\begin{subequations}
\begin{align}
    \hspace*{-0.3cm}\frac{1-N-r_{\ast}N'}{2} \left[1 + \frac{2\lstar^{2}}{r_{\ast}}(1-3N) \vp'  f_{,\vp} \right]
- \frac{r_{\ast}^2}{2} N {\vp'}^2
-\lstar^{2}
(N - 1)\left\{ 2 N {\vp'}^{2} f_{,\vp\vp} +
\left[(1-3N)\frac{\vp'}{r_{\ast}} + 2 N \vp''\right]  f_{,\vp}
\right\} &= 0 \,,
\label{eq:dndr}
\\
\frac{\sigma'}{\sigma}\left[ 1 + \frac{2\lstar^{2}}{r_{\ast}} (1-3N) \vp' f_{,\vp} \right]
- \left[
r_{\ast} {\vp'}^2 - \frac{2\lstar^{2}}{r_{\ast}} (N-1) \left( {\vp'}^2 f_{,\vp\vp} + \vp'' f_{,\vp} \right)
\right] &= 0\,,
\label{eq:dsigmadr}
\\
\vp'' \left[
1
+ \frac{2 \lstar^{2}}{r_{\ast}} (1 - 7N) \vp' f_{,\vp}
- \frac{12 \lstar^{4}}{r_{\ast}^4} \left[ (N-1)^2 + 2 r_{\ast}^2 (1 - 3N)N {\vp'}^2 \right] {f_{,\vp}}^2
\right.
\nonumber \\
\left.
+ \frac{8 \lstar^{6}}{r_{\ast}^5} \left\{ 6(N-1)^2 + [1 + 3(2-5N)N] r_{\ast}^2 {\vp'}^2  \right\} N \vp' {f_{,\vp}}^3
\right]
\nonumber \\
+ \lstar^{2} \left\{
\frac{f_{,\vp}}{r_{\ast}^4 N} \left[ 3(1-N)^2 + 2 r_{\ast}^2 \left(1 - N - 12 N^2 \right) {\vp'}^2  - N r_{\ast}^4 {\vp'}^4 \right]
-\frac{2}{r_{\ast}} (1-N) {\vp'}^{3} f_{,\vp\vp}
\right\}
\nonumber \\
+ \frac{4 \lstar^{4}}{r_{\ast}^5} \left\{
\left[ - 3(N-1)^2 + 2 r_{\ast}^2 (-1 + 3N)(-1+7N) {\vp'}^2 + N r_{\ast}^4 {\vp'}^4  \right] f_{,\vp}
\right.
\nonumber \\
\left. - r_{\ast} (-1+N) \left[ 3 (-1+N) + r_{\ast}^2 (-1+3N) {\vp'}^2 \right] \vp' \, f_{,\vp\vp}
\right\} \vp' \, f_{,\vp}
\nonumber \\
\frac{8 \lstar^{6}}{r_{\ast}^5} \left[
r_{\ast} (1 - 3N)^2 (1 - 5N) \vp' f_{,\vp} + 3 N (N-1)^2 (2 + r_{\ast}^2 {\vp'}^2) f_{,\vp\vp}
\right] {\vp'}^3 \, {f_{,\vp}}^2
+ \frac{1+N}{r_{\ast}N}\vp' &=0\,.
\label{eq:ddsigmaddr}
\end{align}
\label{eq:eom_solve_num}
\end{subequations}
\end{widetext}
This is a system of three coupled ordinary differential equations for $N'$, $\sigma'$ and $\vp''$, which then requires four initial conditions.
The system can be solved numerically once a particular function $f$ and
value of $\lstar$ have been chosen.
For example, choosing $f = 2 \vp$, we recover Eqs.~(3.3)-(3.4) from
Ref.~\cite{Delgado:2020rev} (see also~\cite{Yunes:2011we}).

To obtain black hole solutions, we now identify $r_H$ with the horizon radius and assume that the functions $N$, $\sigma$ and $\vp$ admit power series
expansions near $r_\ast=1$ as:
\begin{subequations}
\label{eq:horizon_exp}
\begin{align}
N &= N^{H}_1 (r_{\ast} - 1) + \dots \,,
\\
\sigma &= \sigma_H + \sigma_1^H(r_{\ast} - 1) + \dots \,,
\\
\vp &= \vp_H + \vp^{H}_1 (r_{\ast} - 1) + \dots\,.
\end{align}
\end{subequations}
We can substitute these expressions into Eqs.~\eqref{eq:eom_solve_num} and solve
order-by-order to fix all their coefficients in terms of $\lstar$,
$\varphi_H$ and $\sigma_H$ only. In particular, we find
\begin{align}
\vp^{H}_1 &= - \frac{1 - \sqrt{1 - 24\, \lstar^4 \, f_{,\vp}(\vp_H)^2}}{4 \lstar^2 f_{,\vp}(\vp_H)}\,,
\label{eq:horizon_coeffs}
\end{align}
from which we conclude that $\lstar$ and $\varphi_H$ must satisfy the well-known condition~\cite{Antoniou:2017acq}
\begin{equation}
24\, \lstar^4\, f_{,\vp}(\vp_H)^2 < 1
\label{eq:scalarh_bound_no_sw}
\end{equation}
for $\vp'$ to be real at the horizon, hence restricting the range of allowed values of
$\vp_H$ given $\lstar=\ell/r_H$.

We numerically integrate Eqs.~\eqref{eq:eom_solve_num} to find
$N(r_\ast)$, $\sigma(r_\ast)$ and $\varphi(r_\ast)$ given four initial
conditions on the horizon:
\begin{equation}
N=0, \quad \sigma=\sigma_H, \quad \varphi=\varphi_H, \quad \textrm{and} \quad
\varphi'=\varphi_1^H\,,\label{eq:fourInitialConditions}
\end{equation}
where $\varphi_1^H$ is given by Eq.~\eqref{eq:horizon_coeffs}.
Note that the numerical value of $\sigma_H$ is pure gauge: it can always be
absorbed in a rescaling of time $t$, cf.~Eq.~\eqref{eq:line_element}.
Hence, black hole solutions depend on two integration constants only,
$\varphi_H$ and $\lstar=\ell / r_H$.
The latter fully takes into account the dependence on the fundamental coupling $\ell$, which only enters through this ratio in Eqs.~\eqref{eq:eom_solve_num}.

We can also expand $N$, $\sigma$ and $\vp$ in inverse powers of $r_{\ast}$ to
study their asymptotic behavior at spatial infinity, i.e. for $r_{\ast} \gg 1$.
By substituting the series
\begin{subequations}
\begin{align}
N &= 1 - \frac{2M_{\ast}}{r_{\ast}} + \frac{N^{\infty}_2}{r_{\ast}^2} + \frac{N^{\infty}_3}{r_{\ast}^3} + \dots\,,
\label{eq:asympt_expansion_n}
\\
\sigma &= 1 + \frac{\sigma^{\infty}_1}{r_{\ast}} + \frac{\sigma^{\infty}_2}{r_{\ast}^2} + \frac{\sigma^{\infty}_3}{r_{\ast}^3} + \dots\,,
\label{eq:asympt_expansion_sigma}
\\
\vp &= \vpbg + \frac{Q_{\ast}}{r_{\ast}} + \frac{\vp^{\infty}_2}{r_{\ast}^2} + \frac{\vp^{\infty}_3}{r_{\ast}^3} + \dots\,,
\label{eq:asympt_expansion_vp}
\end{align}
\label{eq:asympt_expansions}%
\end{subequations}
into Eqs.~\eqref{eq:eom_solve_num} and solving iteratively we find
\begin{align}
    N^{\infty}_2 &= Q_{\ast}^2\,, \quad  N^{\infty}_3 =  M_{\ast} Q_{\ast}^2\,,
\nonumber \\
    \sigma^{\infty}_1 &= 0\,, \quad  \sigma^{\infty}_2 = - Q_{\ast}^2 / 2\,, \quad \sigma^{\infty}_3 = - 4 M_{\ast} Q_{\ast}^2 / 3\,,
\nonumber \\
    \vp^{\infty}_2 &= M_{\ast} Q_{\ast} \,, \quad  \vp^{\infty}_3 = (8 M_{\ast}^2 Q_{\ast} - Q_{\ast}^3) / 6\,.
\end{align}
At all orders, the coefficients entering Eq.~\eqref{eq:asympt_expansions} are
functions of \emph{three} constants $M_{\ast}$, $Q_{\ast}$ and $\vpbg$.
But for black holes, all three quantities are fixed by the \emph{two} integration constants
$\lstar$ and $\varphi_H$ [once $\sigma_H$ is set to ensure the gauge
$\sigma=1$ at infinity as in Eq.~\eqref{eq:asympt_expansion_sigma}]: the ``scalar hair'' is said to be ``secondary''~\cite{Coleman:1991ku,R:2022cwe}.
Although $M_{\ast}$, $Q_{\ast}$ and $\vpbg$ can be obtained from the $\mathcal O(r^{-1}_\ast)$
fall-off of $N$ and $\varphi$, we also use subleading terms up to
$\mathcal O(r^{-3}_\ast)$ and the expansion of $\sigma$ to accurately
extract them from our numerical integration, which terminates at a
finite $r_{\ast}$.

Let us conclude this section by further illustrating the consequences of
Eq.~\eqref{eq:scalarh_bound_no_sw}, as it will play an important role below.
Solving analytically for the coefficients of Eqs.~\eqref{eq:horizon_exp} up to $\varphi_4^H$, $N_4^H$ and $\sigma_3^H$, we can compute the near-horizon scalar field and Gauss-Bonnet invariant \eqref{eq:gb_for_line_element} as
\begin{subequations}
\begin{align}
\vp &=\vp_{H}+ \sum_{n=1}^{4}\vp^{H}_n (r_{\ast} - 1)^n + \mathcal O(r_\ast-1)^5\,,\label{nearHorizonPhi} \\
\sG \, r_H^4&=g_H+\sum_{n=1}^{2}g^H_n (r_{\ast} - 1)^n + \mathcal O(r_\ast-1)^3\,,\label{nearHorizonGB}
\end{align}\label{nearHorizonPhiAndGB}%
\end{subequations}
where $\varphi_1^H$ is given by Eq.~\eqref{eq:horizon_coeffs} and where
the other coefficients are long functions of $\lstar$ and $\varphi_H$ [but not of the gauge-fixing quantity $\sigma_H$, see below Eq.~\eqref{eq:fourInitialConditions}] available online~\cite{FLJRepo}.
However, near the saturation of the bound \eqref{eq:scalarh_bound_no_sw}, i.e., when $\epsilon^2=1-24\,\lstar^{4}\, f_{,\vp}(\vp_H)^2\ll 1$ but $\epsilon \neq 0$, we find
\begin{subequations}
\begin{align}
\vp^{H}_1 &= -\sqrt{\frac{3}{2}}+\mathcal O(\epsilon)\,, \\
\vp^{H}_2 &= \sqrt{\frac{3}{2}}\frac{9}{16}\frac{\chi}{\epsilon}+\mathcal O(\epsilon^0)\,,\\
\vp^{H}_3 &= -\sqrt{\frac{3}{2}}\frac{27}{128}\frac{\chi^2}{\epsilon^3}+\mathcal O(\epsilon^{-2})\,, \\
\vp^{H}_4 &= \sqrt{\frac{3}{2}}\frac{729}{4096}\frac{\chi^3}{\epsilon^5}+\mathcal O(\epsilon^{-4})\,,
\end{align}\label{eq:phiHorizonCoeffs}%
\end{subequations}
and
\begin{subequations}
\begin{align}
g_H&=48+\mathcal O(\epsilon)\,,\label{eq:GBhorizonCoeff1}\\
g^{H}_1&=-216 \, \frac{\chi}{\epsilon}+\mathcal O(\epsilon^{0})\,, \\
g^{H}_2&=\frac{729}{4}\frac{\chi^2}{\epsilon^3}+\mathcal O(\epsilon^{-2})\,,
\end{align}\label{eq:GBhorizonCoeffs}%
\end{subequations}
with $\chi=3+4 \lstar^{2} f_{,\vp\vp}(\vp_H)$. While $\varphi_1^H$
and $g_H$ are finite and do not depend on $f(\varphi)$ in this limit,
every other coefficient in Eqs.~\eqref{eq:phiHorizonCoeffs}-\eqref{eq:GBhorizonCoeffs} is singular.
We find qualitatively similar results, that we report in Appendix~\ref{app:RicciKretschmann}, for the Ricci and Kretschmann curvature invariants $R$ and $\mathcal{K}=R^{\mu\nu\rho\sigma}R_{\mu\nu\rho\sigma}$. In Sec.~\ref{sec:coup_ssym}, we will compare the analytic predictions \eqref{nearHorizonPhiAndGB} to numerical results.

\section{Black hole sensitivities}
\label{sec:bh_sens}

The PN dynamics of black hole binaries in ESGB gravity was
studied in
Refs.~\cite{Yagi:2011xp,Julie:2019sab,Shiralilou:2020gah,Shiralilou:2021mfl} in
the weak-field, slow orbital velocity limit.
In this context, Refs.~\cite{Julie:2019sab,Cardenas:2017chu} showed that when
finite-size corrections (e.g.~tidal and out-of-equilibrium effects)  can be neglected, each black
hole is described by a sequence of static configurations with identical Wald
entropy $\sS_{\rm W}$ defined as~\cite{Wald:1993nt,Iyer:1994ys,Torii:1996yi}
\begin{equation}
\sS_{\rm W} = \frac{\sA_{H}}{4} + 4 \pi \lsqr f(\vp_{H})\,,
\label{eq:wald}
\end{equation}
where $\sA_{H}$ is the horizon surface area (here $4 \pi r_H^2$).
The PN Lagrangian \cite{Julie:2019sab} and fluxes
\cite{Yagi:2011xp,Shiralilou:2020gah,Shiralilou:2021mfl} then depend on
``sensitivities'' which characterize the response of each black hole to its
adiabatically changing scalar-field environment.

More precisely, the sensitivity of a black hole is defined as the
logarithmic change in $M$ with respect to $\vpbg$ [cf. Eq.~\eqref{eq:asympt_expansion_vp}] at \emph{fixed Wald entropy
$\sS_{\rm W}$} \cite{Julie:2019sab,Cardenas:2017chu}:
\begin{align}
\alpha \equiv \left. \frac{\dd \ln M}{\dd \vpbg} \right\vert_{\sS_{\rm W}}
=\frac{1}{M}\left.\frac{\dd M}{\dd \vpbg}\right\vert_{\sS_{\rm W}}\,,
\label{eq:sensitivity}
\end{align}
and we denote its derivative with respect to $\vpbg$, which also enters the
1PN Lagrangian~\cite{Julie:2019sab}, by $\beta$:
\begin{align}
\beta\equiv\left.\frac{\dd \alpha}{\dd\vpbg}\right\vert_{\sS_{\rm W}}\,.\label{eq:sensitivity_beta}
\end{align}
Equation~\eqref{eq:sensitivity} is similar to the notion of
sensitivity for self-gravitating bodies (such as neutron stars) in
scalar-tensor theories, defined as the logarithmic change in the ADM mass $M$
with respect to some external $\vpbg$, but at fixed baryonic
mass~\cite{Damour:1992we,Damour:1995kt,Damour:1996ke}.

In a binary, the sensitivity $\alpha$ of a body accounts for the
readjustments of its ADM mass $M$ and scalar charge $Q$ during the inspiral.
This sensitivity has to be evaluated at a value $\vpbg$ corresponding to its time-varying but spatially homogeneous background
scalar field sourced by the far-away companion (recall that finite-size effects are here neglected).
For our purposes, we take $\vpbg$ to be just some nonzero scalar
field value in which the isolated black hole is embedded.

Reference~\cite{Julie:2019sab} also showed that the variation of $\sS_{\rm W}$, $M$
and $\vpbg$ with respect to the black hole's integration constants
(here $\lstar$ and $\varphi_H$, see Sec.~\ref{subsec:nonrotating_bhs})
must satisfy the identity
\begin{equation}
T \, \delta \sS_{\rm W}=\delta M+Q\,\delta\vpbg\,,
\label{eq:first_law}
\end{equation}
where $T$ is the temperature~\cite{Julie:2019sab}, whose expression we do not need here.
Comparing this first law of thermodynamics in the case of interest ($\delta\sS_{\rm
W}=0$) with the definition \eqref{eq:sensitivity} we get
\begin{equation}
    \alpha= - Q / M \,,
    \label{eq:sensitivity_fun_charge}
\end{equation}
which provides a second, independent way of calculating the black hole sensitivity.

We numerically calculate the sensitivity $\alpha$ as follows:
\begin{enumerate}
\item\label{itm:step1} Fix a value of the dimensionless ratio $\ell / \mu$, where
\begin{equation}
\mu^2 =
\frac{\sS_{\rm W}}{4\pi}
\label{eq:irrmass}
\end{equation}
is the irreducible mass squared~\cite{Christodoulou:1970wf}. From
Eq.~\eqref{eq:wald}, this ratio is related to $\lstar=\ell/r_H$ and $\varphi_H$ through
\begin{equation}
(\ell / \mu)^{-2} = (4 \lstar^2)^{-1} + f(\vp_H)\,.
\label{eq:a_mu_ratio}
\end{equation}
\item\label{itm:step2} Choose a value of the scalar field at the horizon $\vp_H$.
\item\label{itm:step3} Using Eq.~\eqref{eq:a_mu_ratio}, solve for $\lstar$ and use this
value to numerically construct a black hole, integrating the system \eqref{eq:eom_solve_num} with initial conditions~\eqref{eq:fourInitialConditions} at $r_{\ast} = 1$
up to a large value of $r_{\ast}$. The constant $\sigma_H$ is pure gauge and we fix it by requiring
that the line element~\eqref{eq:line_element} asymptotes to $\sigma=1$.
\item\label{itm:step4} Calculate the quantities
$\vpbg$, $M_{\ast}$ and $Q_{\ast}$
from the asymptotic expressions~\eqref{eq:asympt_expansions}.
\item\label{itm:step5}
Repeat steps~(\ref{itm:step2}) to~(\ref{itm:step4}) for the
range of $\vp_H$ values allowed by Eq.~\eqref{eq:scalarh_bound_no_sw},
hence obtaining a family of constant Wald entropy black holes.
For such a family the condition \eqref{eq:scalarh_bound_no_sw} becomes
\begin{equation}
\tfrac{3}{2} f_{,\vp}(\vp_H)^2 < [(\mu / \ell)^2 - f(\vp_H)]^2\,.
\label{eq:scalarh_bound_sw}
\end{equation}
\item\label{itm:step6} Since a constant $\sS_{\rm W}$ is equivalent to a constant $\mu$, we can
calculate $\alpha$ numerically by inserting
\begin{align}
M / \mu &= M_{\ast} (\ell / \mu) \lstar^{-1}
\label{eq:admm_over_irrm}
\end{align}
into Eq.~\eqref{eq:sensitivity}, or by directly computing the
scalar-charge-to-mass ratio $- Q_{\ast}/M_{\ast} = - Q/M$ [cf.
Eq.~\eqref{eq:sensitivity_fun_charge}], which is invariant under
rescaling by $r_{H}$.
Once we know $\alpha$, we calculate $\beta$ using Eq.~\eqref{eq:sensitivity_beta}.
\end{enumerate}

The numerical methods used in this paper are summarized in Appendix~\ref{app:numerical_methods}. In calculations that will follow, we will be interested in the behavior of certain quantities close to the saturation of Eq.~\eqref{eq:scalarh_bound_sw}. Numerically, we can only reach a minimum value of $\epsilon = |\vp_H - \vp_H^\mx|$, where $\vp_H^\mx$ saturates Eq.~\eqref{eq:scalarh_bound_sw}. Here we take $\epsilon\sim 10^{-5}$, and we will refer to the limiting process as ``approaching the saturation of Eq.~\eqref{eq:scalarh_bound_sw}.''

In the context of PN calculations, $M / \mu $, $\alpha$ and $\beta$ must be viewed
as functions of the asymptotic scalar field $\vpbg$, the irreducible
mass $\mu$, and the fundamental constant $\ell$.
The last two only contribute through their dimensionless ratio
$\ell / \mu$, since the only free parameter entering the differential
equations~\eqref{eq:eom_solve_num} is $\lstar$, which is in turn related to
$\ell / \mu$ through Eq.~\eqref{eq:a_mu_ratio}.
We find full agreement between both methods specified in step~\ref{itm:step6} above to
compute the sensitivity $\alpha$.
This proves that our families of constant-entropy black holes are consistent
with the first law of thermodynamics: see the discussion following Eq.~\eqref{eq:first_law}.

When possible, our numerical results will be compared against the analytical black hole sensitivities obtained in a small-$\ell/\mu$ expansion around Schwarzschild in
Ref.~\cite{Julie:2019sab}.
The results there have the schematic form
\begin{equation}
    \alpha = - \frac{x}{2} - \sum_{n = 2}^{N} A_{n}(\vpbg)\, x^{n} \,,
    \quad
    x = \frac{\lsqr f'(\vpbg)}{\mu^2}\,,
    \label{eq:analytical_sens}
\end{equation}
where the coefficients $A_{n}$ depend on $f$ and its derivatives evaluated at $\vpbg$.
The calculation in Ref.~\cite{Julie:2019sab} obtained the
series~\eqref{eq:analytical_sens} up to $N=4$ and here we extend it up
to $N=10$ for a more careful comparison with our numerical results.
These lengthy results are available online~\cite{FLJRepo}.

In the following subsections we compare the numerical and analytical
calculations for black holes for three particular choices of the coupling
function $f$.

\subsection{Shift-symmetric theory}
\label{sec:coup_ssym}

As a first example, consider the theory
\begin{equation}
    f(\vp) = 2 \vp\,,\label{eq:def_shift_sym_theory}
\end{equation}
such that the action \eqref{eq:action} becomes invariant under the shift $\vp \to \vp + \Delta\vp$, where $\Delta\vp$ is a constant.
The condition for the existence of a real scalar field at the horizon of constant entropy black holes \eqref{eq:scalarh_bound_sw} simplifies to
\begin{equation}
\vp_{H} < \frac{1}{2} \left(\frac{\mu^2}{\lsqr} - \sqrt{6}\right)\,.
\label{eq:ss_max_scalar}
\end{equation}

\begin{figure}[t]
\includegraphics[width=\columnwidth]{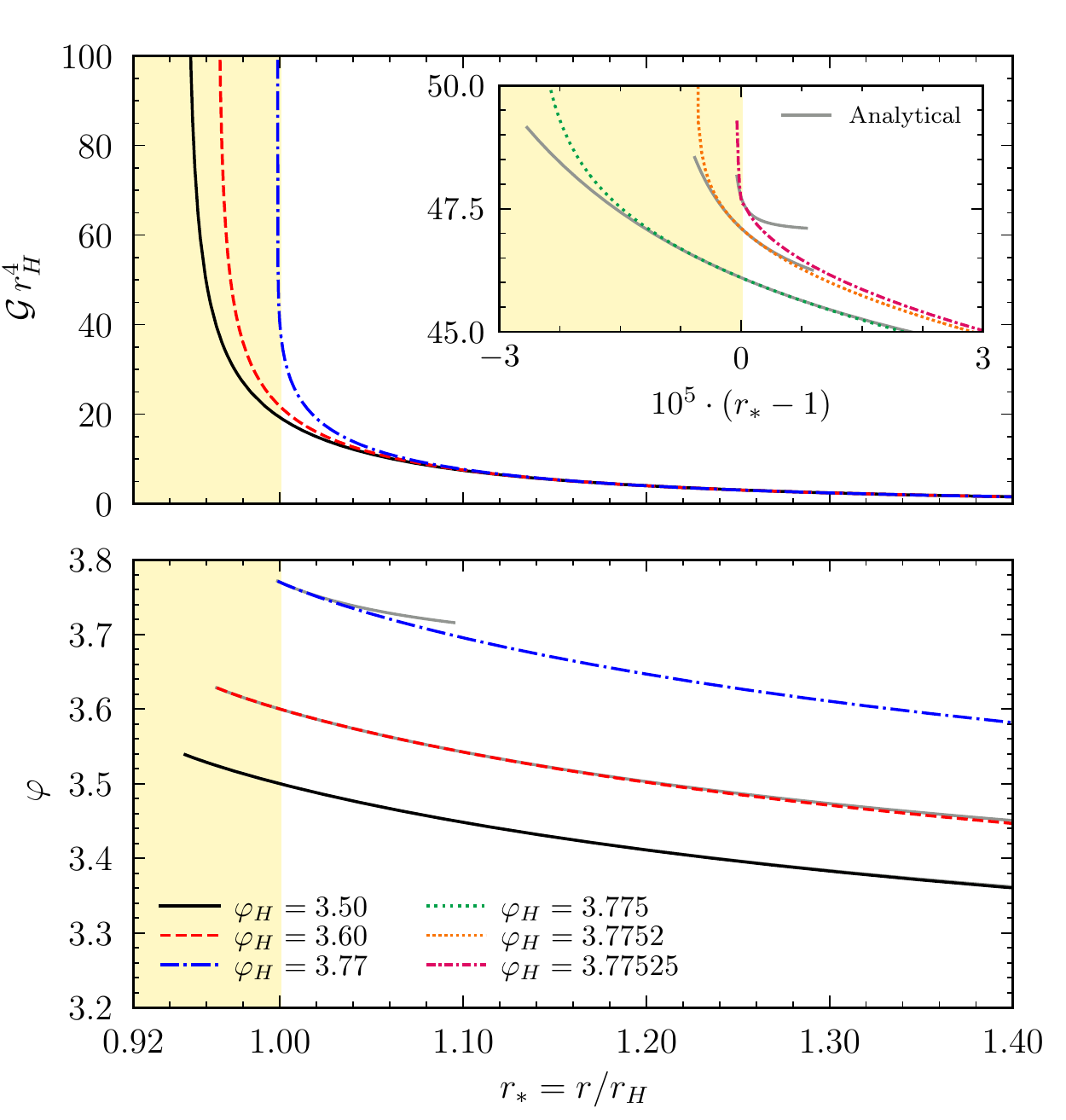}
\caption{
A sequence of constant $(\ell / \mu)^2 = 10^{-1}$ black holes
with $\vp_{H}$ approaching $\vp_{H}^{\mx}\approx 3.77526$, inside
(shaded region, $r_{\ast} < 1$) and outside ($r_{\ast} > 1$) the horizon.
Top panel: the Gauss-Bonnet scalar diverges at $r_{\ast} = \{ 0.956, 0.965, 0.999 \}$
when $\varphi_H= \{ 3.50, 3.60, 3.77 \}$ approaches $\varphi_H^{\mx}$.
The inset also shows $\vp_{H}=\{3.775, 3.7752, 3.77525\}$,
closing in into $\vp_{H}^{\mx}$ one order of magnitude after another.
Bottom panel: the scalar field is finite at the curvature singularity.
In the bottom panel and in the top panel's inset, the numerical results agree at $r_\ast=1$ with the $(2,2)$-Pad\'e resummation of Eq.~\eqref{nearHorizonPhi} and the $(1,1)$-Pad\'e resummation of Eq.~\eqref{nearHorizonGB}, respectively.
In particular, $\mathcal{G} \, r^{4}_{H}$ approaches the value $48$ when $\vp_{H}$ approaches $\vp_{H}^{\mx}$, hence recovering Eq.~\eqref{eq:GBhorizonCoeff1},
with $\mathcal{G} \, r^{4}_{H} \approx 47.73$ for $\varphi_{H} = 3.77525$. Both $\sG \, r_H^4$ and $\varphi$ converge to finite values for all $r_\ast\geqslant1$ when $\vp_{H}$ is increased towards $\vp_{H}^{\mx}$.
}
\label{fig:ss_sensitivity_vs_phi0_bh_props}
\end{figure}

\begin{figure}
\includegraphics[width=\columnwidth]{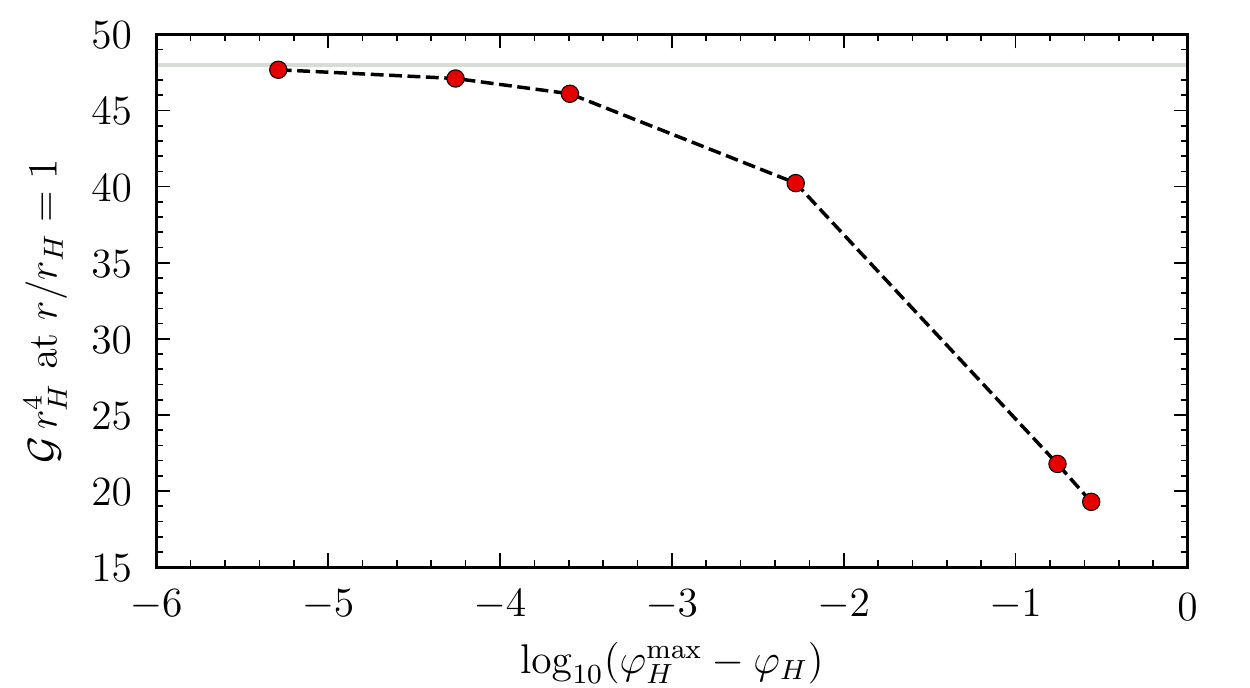}
\caption{Numerical Gauss-Bonnet scalar  $\sG \, r_{H}^{4}$
evaluated at the event horizon $r_{\ast} = r / r_{H} = 1$ for a sequence of black holes with constant
$(\ell / \mu)^2 = 10^{-1}$.
As the scalar field at the horizon $\vp_{H}$ approaches its maximum allowed value $\vp^{\mx}_{H} \approx 3.77526$, the Gauss-Bonnet scalar tends to the limit $\sG \, r_{H}^{4}=48$ (horizontal line) predicted analytically by Eq.~\eqref{eq:GBhorizonCoeff1}.}
\label{fig:gb_at_horizon}
\end{figure}

\begin{figure*}[ht]
\includegraphics[width=.95\columnwidth]{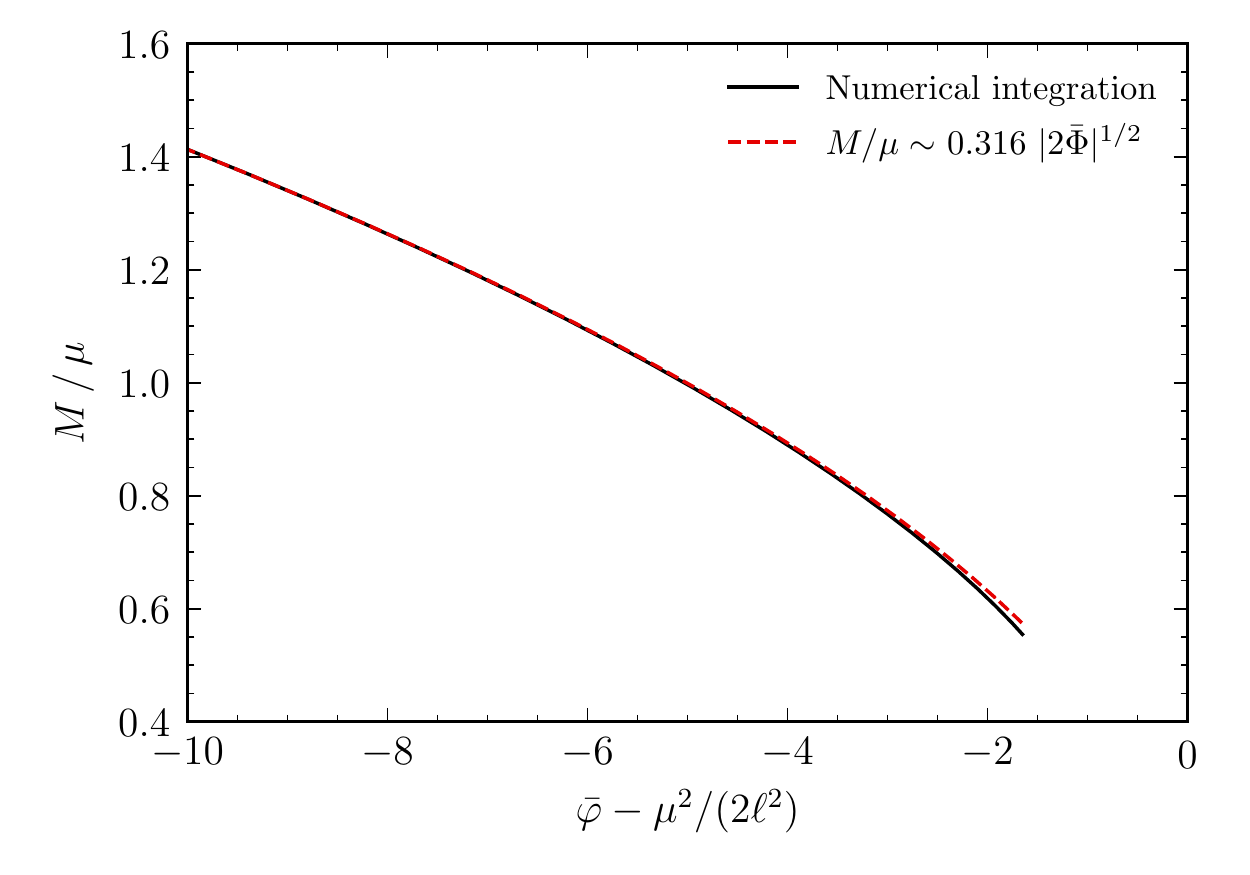}
\includegraphics[width=.95\columnwidth]{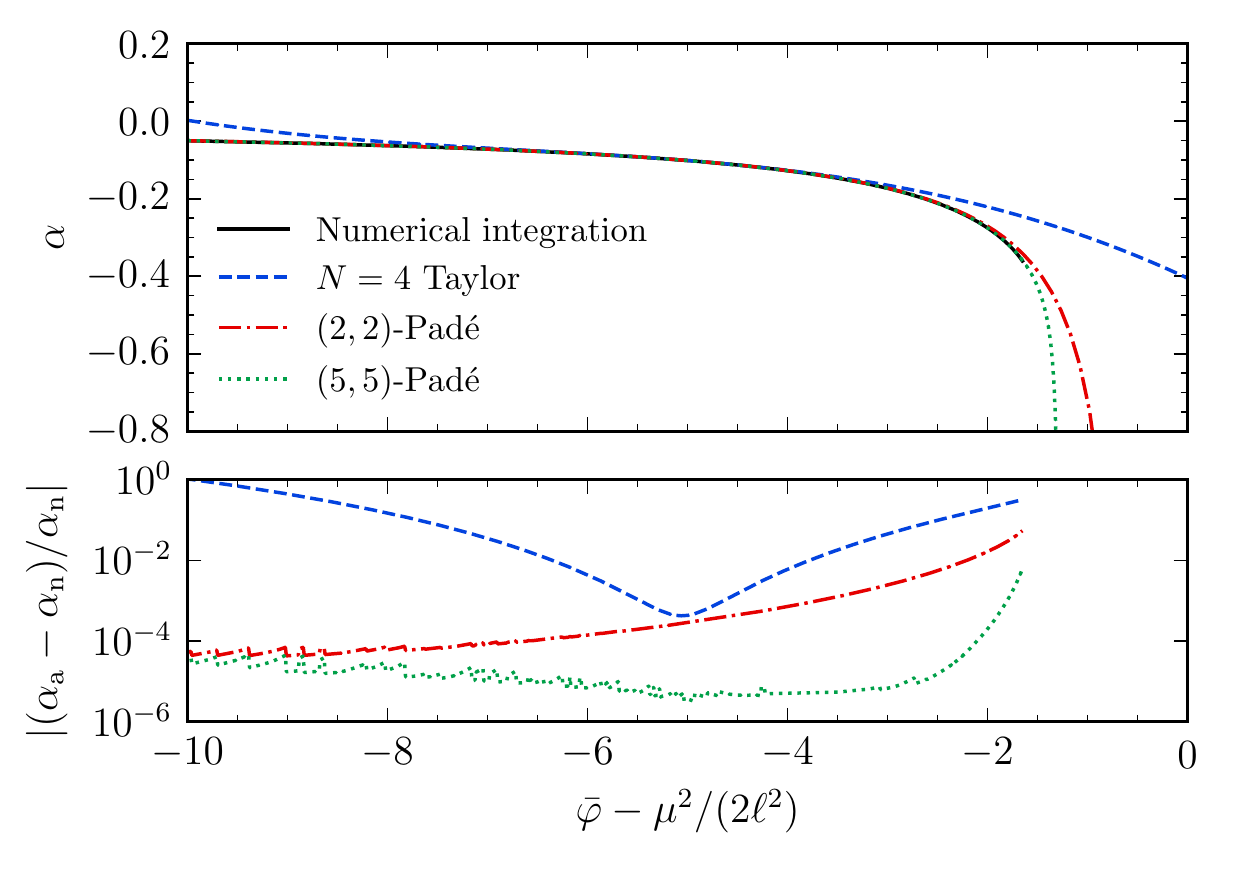}
\caption{
Black hole mass and sensitivity in the shift-symmetric theory \eqref{eq:def_shift_sym_theory} as
functions of the quantity $\vpbg - \mu^{2} / (2\lsqr)$ introduced in Eq.~\eqref{eq:combinationShiftSym}.
Left panel: the numerical ADM-to-irreducible mass ratio $M/\mu$.
Right panel: the numerical sensitivity $\alpha$ and its analytic estimates from Eq.~\eqref{eq:analytical_sens} with $N=4$, its $(2,2)$-Pad\'e resummation, and the $(5,5)$-Pad\'e resummation of Eq.~\eqref{eq:analytical_sens}
with $N=10$.
The bottom-right panel shows the fractional error between analytic (``a'') and numerical (``n'') calculations.
The numerical sensitivity and its ($5,5$)-Pad\'e counterpart show excellent agreement, modulo one substantial qualitative difference: the Pad\'e approximant is singular as an artifact of the method, while the numerical sensitivity curve ends at $\vpbg - \mu^{2} / (2\lsqr) \approx -1.651$ as we approach saturation of Eq.~\eqref{eq:ss_max_scalar}.
In the limit $\bar\Phi=\vpbg - \mu^{2} / (2\lsqr)\to-\infty$, we have $M/\mu=A|2\bar\Phi|^{1/2}+\mathcal O|\bar\Phi|^{-1/2}$ with $A\approx 0.316$ and $\alpha\to 0$, and at the end points we find $M/\mu\approx 0.555$ and $\alpha \approx - 0.350$.
}
\label{fig:ss_sensitivity_vs_phi0}
\end{figure*}

In Fig.~\ref{fig:ss_sensitivity_vs_phi0_bh_props}, we show the radial profiles of the Gauss-Bonnet invariant (top panel)
and scalar field (bottom panel), both inside (shaded region, $r_{\ast} < 1$) and outside ($r_{\ast} > 1$)
the horizon, for a sequence of constant $(\ell / \mu)^2 = 10^{-1}$ black holes, as $\vp_{H}$ approaches the maximum value $\vp_{H}^{\mx}=(10-\sqrt{6})/2 \approx 3.77526$ saturating Eq.~\eqref{eq:ss_max_scalar}.
As is well-known from, e.g., Refs.~\cite{Sotiriou:2014pfa,Sullivan:2019vyi},
the black holes have a hidden curvature singularity which is driven towards
the horizon as $\varphi_H$ approaches $\vp_{H}^{\mx}$.
However, in this paper we wish to shed new light on this phenomenon.
To this aim, we carefully let $\vp_{H}$ approach $\vp_{H}^{\mx}$ one order of
magnitude after another, since we cannot saturate
Eq.~\eqref{eq:ss_max_scalar} exactly due to the finite precision of numerical
integrations.

Figure~\ref{fig:ss_sensitivity_vs_phi0_bh_props} shows the excellent agreement
at $r_\ast=1$ between the numerical profiles and their analytic near-horizon
counterparts in Eqs.~\eqref{nearHorizonPhiAndGB}.
Moreover, a striking feature of the radial profiles of $\sG r_H^4$ and
$\varphi$ is that they both converge to finite values for all $r_\ast\geqslant1$ when $\vp_{H}$
is increased towards $\vp_{H}^{\mx}$ despite, meanwhile, the hidden
curvature singularity approaching the horizon.
Our results provide numerical evidence that as $\varphi_H$ approaches $\vp_{H}^{\mx}$, the Gauss-Bonnet scalar
reaches the finite value $\sG r_H^4=48$ as $r_\ast\to 1$ with $r_\ast> 1$, as shown in Fig.~\ref{fig:gb_at_horizon}. This value of $\sG r_H^4=48$ coincides with the first term in the analytic, theory-independent prediction of Eq.~\eqref{eq:GBhorizonCoeff1}.
Given these results, it is not clear that a naked singularity arises when one saturates the bound in Eq.~\eqref{eq:scalarh_bound_no_sw} (see e.g.~\cite{Sotiriou:2014pfa,Sullivan:2019vyi}).
We find qualitatively similar behavior for the Gauss-Bonnet scalar and scalar
field in the other ESGB theories considered in this paper.
We also provide analytical near-horizon expansions of the Ricci and Kretschmann scalars valid for all ESGB theories in Appendix~\ref{app:RicciKretschmann}.

Let us now return to the sensitivity $\alpha$. As explained below
Eq.~\eqref{eq:admm_over_irrm}, this quantity must be seen as a function of
$\vpbg$ and of the ratio $\ell/\mu$. However, we can exploit the theory's shift
symmetry to calculate it for all values of $\ell/\mu$ at once.
Indeed, the Wald entropy \eqref{eq:wald} now reads
\begin{equation}
\sS_{\rm W} =\pi (r_H^2 + 8 \lsqr \vp_H) \,,
\end{equation}
and it is linear in $\varphi_H$.
The sensitivity can therefore only depend on the combination
\begin{equation}
\vpbg - \frac{\mu^2}{2\lsqr}\label{eq:combinationShiftSym}
\end{equation}
which is invariant under a scalar field shift, i.e., under the simultaneous redefinitions
$ \vpbg \to \vpbg + \Delta \vp$
and
$\mu^2=\sS_{\rm W}/4\pi  \to \mu^2 + 2 \lsqr \Delta \vp$.
This means that the sensitivities of black holes with constant irreducible masses $\mu_{A}$ and $\mu_{B}$, in shift-symmetric theories with fundamental couplings $\ell_{A}$ and $\ell_{B}$ respectively, are related to each other as $ \alpha_{A}(\vpbg) = \alpha_{B} (\vpbg + \Delta \vpbg)$ with
\begin{align}
\quad \Delta \vpbg &= \frac{1}{2}
                 \left( \frac{\mu^2_{B}}{\ell_{B}^{2}} - \frac{\mu^2_{A}}{\ell_{A}^{2}} \right)\,.
    \label{eq:delta_vp_inf_ss}
 \end{align}

That this is the case was verified in the perturbative calculation of
Ref.~\cite{Julie:2019sab}, but it can also be proven nonperturbatively as follows.
Substitute $\vp=\Phi+\mu^2/(2\lsqr)$ into the differential system~\eqref{eq:eom_solve_num} with $\lstar^2=-1/(8\Phi_H)$ [cf.
Eq.~\eqref{eq:a_mu_ratio}], and observe that the result depends on a single
parameter, $\Phi_H=\varphi_H-\mu^2/(2\lsqr)$.
Then, integrate the system using the initial conditions \eqref{eq:fourInitialConditions} on the horizon $r_\ast=1$, i.e. $N(1)=0$, $\Phi(1)=\Phi_H$ and $\Phi'(1)=\Phi_H-(\Phi_H^2-3/2)^{1/2}$, and note that $\Phi_H$ is therefore the only integration constant.
The latter can finally be traded for $\bar\Phi=\vpbg- \mu^2/(2\lsqr)$, which is the asymptotic value of $\Phi$ at large $r_\ast$ that coincides with Eq.~\eqref{eq:combinationShiftSym},
by inverting $\bar\Phi(\Phi_H)$.

In Fig.~\ref{fig:ss_sensitivity_vs_phi0}, we therefore show the ADM-to-irreducible mass ratio $M/\mu$ (left panel) and sensitivity $\alpha$ (right panel) as functions of the combination of Eq.~\eqref{eq:combinationShiftSym}.
The top-right panel also includes analytic approximants of $\alpha$ obtained from the Taylor series \eqref{eq:analytical_sens} with $N=4$, its $(2,2)$-Pad\'e resummation \cite{Julie:2019sab}, and the $(5,5)$-Pad\'e resummation of Eq.~\eqref{eq:analytical_sens} pushed to $N=10$ in this paper. Here $x=2(\ell/\mu)^2$, and we use Pad\'e approximants to accelerate the convergence of our analytic results.
The bottom-right panel shows the relative error between analytic and numerical
calculations.
We relegate a discussion of the quantity $\beta$, deduced from $\alpha$ by means of
Eq.~\eqref{eq:sensitivity_beta}, to Appendix~\ref{app:beta_sensitivity}.

The agreement between the numerical sensitivity and its $(5,5)$-Pad\'e
counterpart is remarkable, modulo one substantial qualitative difference.
The Pad\'e approximants diverge as $\vpbg-\mu^2/(2\lsqr)$ is increased, and
they feature poles as an artifact of the method~\cite{2002nrca.book.....P}.
By comparison, we find
a finite numerical sensitivity,
whose curve terminates earlier than that of the $(5,5)$-Pad\'e approximant, at
\begin{equation}
\vpbg - \frac{\mu^2}{2\lsqr}\lesssim-1.651\ .\label{eq:ss_max_scalar_Infty}
\end{equation}
Indeed, the saturation of this inequality coincides numerically with that of the horizon bound~\eqref{eq:ss_max_scalar}, which, in turn, is related to the
hidden curvature singularity approaching the black hole's horizon, see
Fig.~\ref{fig:ss_sensitivity_vs_phi0_bh_props}.

The role of the scalar background $\vpbg$ of a black hole with fixed Wald
entropy $\sS_{\rm W}=4\pi\mu^2$ is therefore the following:

\begin{enumerate}
\item
when $\vpbg \to -\infty$ the black hole decouples from the scalar field, since
$\alpha=-Q/M$ (as well as its derivatives, such as $\beta$) vanishes.
More precisely, the diagonal ($n,n$)-Pad\'e approximants with
$n\in[ 1,\, 5]$
of Eq.~\eqref{eq:analytical_sens}, which we know up to $N=10$, all predict $\alpha=1/(2\bar\Phi)+\mathcal O(\bar\Phi^{-2})$ with $\bar\Phi=\vpbg-\mu^2/(2\lsqr)\to-\infty$. Integrating Eq.~\eqref{eq:sensitivity} then implies
\begin{equation}
M/\mu=A\,|2\bar\Phi|^{1/2}+\mathcal O|\bar\Phi|^{-1/2}\,,
\end{equation}
which fits our numerical results for $A\approx 0.316$.
We remark that this fit works remarkably well in the whole range of $\bar{\Phi}$
(see Fig.~\ref{fig:ss_sensitivity_vs_phi0}), despite having been obtained
only in the range $\bar{\Phi} \in [-10, \, -9]$;
\item
when $\vpbg$ is increased, the black hole develops a nonzero and negative
sensitivity $\alpha$, and the hidden curvature singularity
approaches the horizon at $\vpbg - \mu^2/(2\lsqr) \approx - 1.651$,
where $M/\mu\approx 0.555$ and $\alpha\approx-0.350$ as shown in
Fig.~\ref{fig:ss_sensitivity_vs_phi0}: see also Eq.~\eqref{eq:ss_max_scalar_Infty}.
\label{itm:ss_ext}
\end{enumerate}
The consequences of point~\ref{itm:ss_ext} above on adiabatically inspiralling black hole
binaries will be investigated in Sec.~\ref{sec:application}.

\subsection{Dilatonic theory}
\label{sec:coup_dilaton}

As a second example, consider the theory
\begin{equation}
f(\vp) = \frac{1}{4} \exp(2 \vp) \,,
\label{eq:def_dilatonic_theory}
\end{equation}
such that the action \eqref{eq:action} is invariant under the simultaneous redefinitions $\vp \to \vp + \Delta\vp$ and $\ell\to\ell\exp(- \Delta\vp)$, where $\Delta\vp$ is a constant.
The condition for the existence of a real scalar field at the horizon of constant entropy black holes \eqref{eq:scalarh_bound_sw} becomes
\begin{equation}
\vp_H + \ln\left( \frac{\ell}{\mu} \right)
<
\frac{1}{2} \ln \left( \frac{4}{1 + \sqrt{6}} \right)
\,.
\label{eq:dl_bound}
\end{equation}

As with the shift-symmetric case, we can exploit the symmetry of the theory to
calculate the sensitivity $\alpha$ for all values of $\ell / \mu$ at once.
Indeed, the Wald entropy \eqref{eq:wald} now reads
\begin{equation}
    \sS_{\rm W} = \frac{1}{4} [r_{H}^2 + 4 \pi \lsqr \exp(2\vp_H)]\,.
\end{equation}
As observed in~\cite{Julie:2019sab}, the sensitivities can therefore only depend on the combination
\begin{equation}
\vpbg +  \ln\left( \frac{\ell}{\mu} \right)\,,
\label{eq:combinationDilaton}
\end{equation}
which is invariant under the simultaneous redefinitions $\vpbg\to\vpbg+\Delta\vp$,
$\vp_{H}\to\vp_{H}+\Delta\vp$ and $\ell\to\ell\exp(-
\Delta\vp)$, since then $\mu=(\sS_{\rm W}/4\pi)^{1/2}$ is also invariant.
In other words, the sensitivities of black holes with constant irreducible
masses $\mu_{A}$ and $\mu_{B}$, in dilatonic theories with fundamental
couplings  $\ell_A$ and $\ell_B$ respectively, are related to each other as
$\alpha_{A}(\vpbg) = \alpha_{B} (\vpbg + \Delta \vpbg)$ with
\begin{equation}
    \Delta \vpbg = \ln
    \left(
        \frac{\ell_{A}/\mu_{A}}{\ell_{B}/\mu_{B}}
    \right)
    \,.
    \label{eq:delta_vp_inf_dil}
\end{equation}

This statement was verified in the perturbative calculation of Ref.~\cite{Julie:2019sab},
but we can again prove it nonperturbatively as follows:
substitute $\vp=\Phi- \ln (\ell/\mu)$ into the differential system
\eqref{eq:eom_solve_num} with $\lstar=(\ell/\mu)/(4-e^{2\Phi_H})^{1/2}$ [cf.
Eq.~\eqref{eq:a_mu_ratio}], and observe that the result only depends on one
parameter, $\Phi_H=\varphi_H+\ln (\ell/\mu)$.
Then, integrate the system using the initial conditions
\eqref{eq:fourInitialConditions} on the horizon $r_\ast=1$, i.e.~$N(1)=0$,
$\Phi(1)=\Phi_H$ and $\Phi'(1)=-x+(x^2-3/2)^{1/2}$ where $x=2e^{-2\Phi_H}-1/2$,
and note that $\Phi_H$ is the only integration constant.
The latter can finally be traded for the asymptotic value of $\Phi$,
$\bar\Phi=\vpbg+\ln (\ell/\mu)$, which coincides with
Eq.~\eqref{eq:combinationDilaton}, by inverting $\bar\Phi(\Phi_H)$.

\begin{figure*}[t]
\includegraphics[width=\columnwidth]{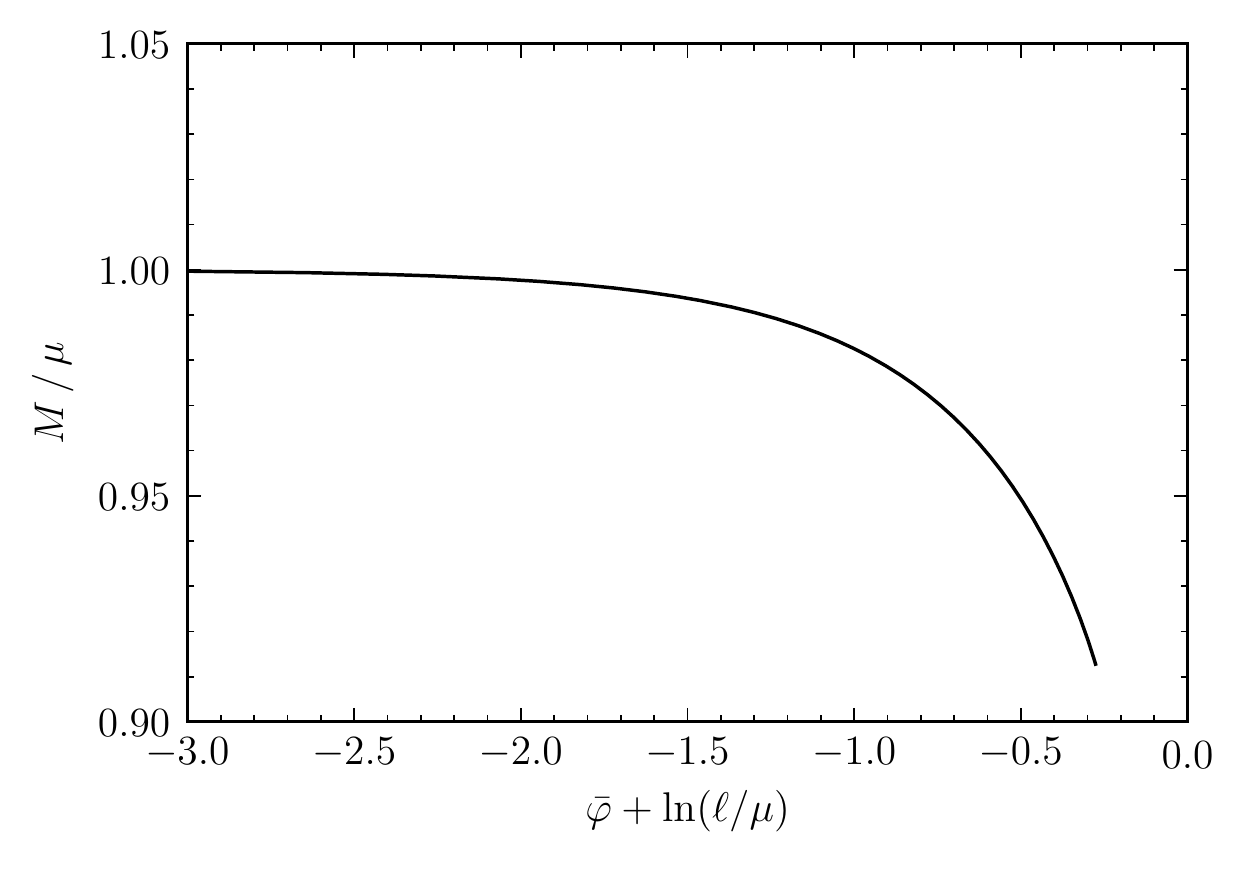}
\includegraphics[width=\columnwidth]{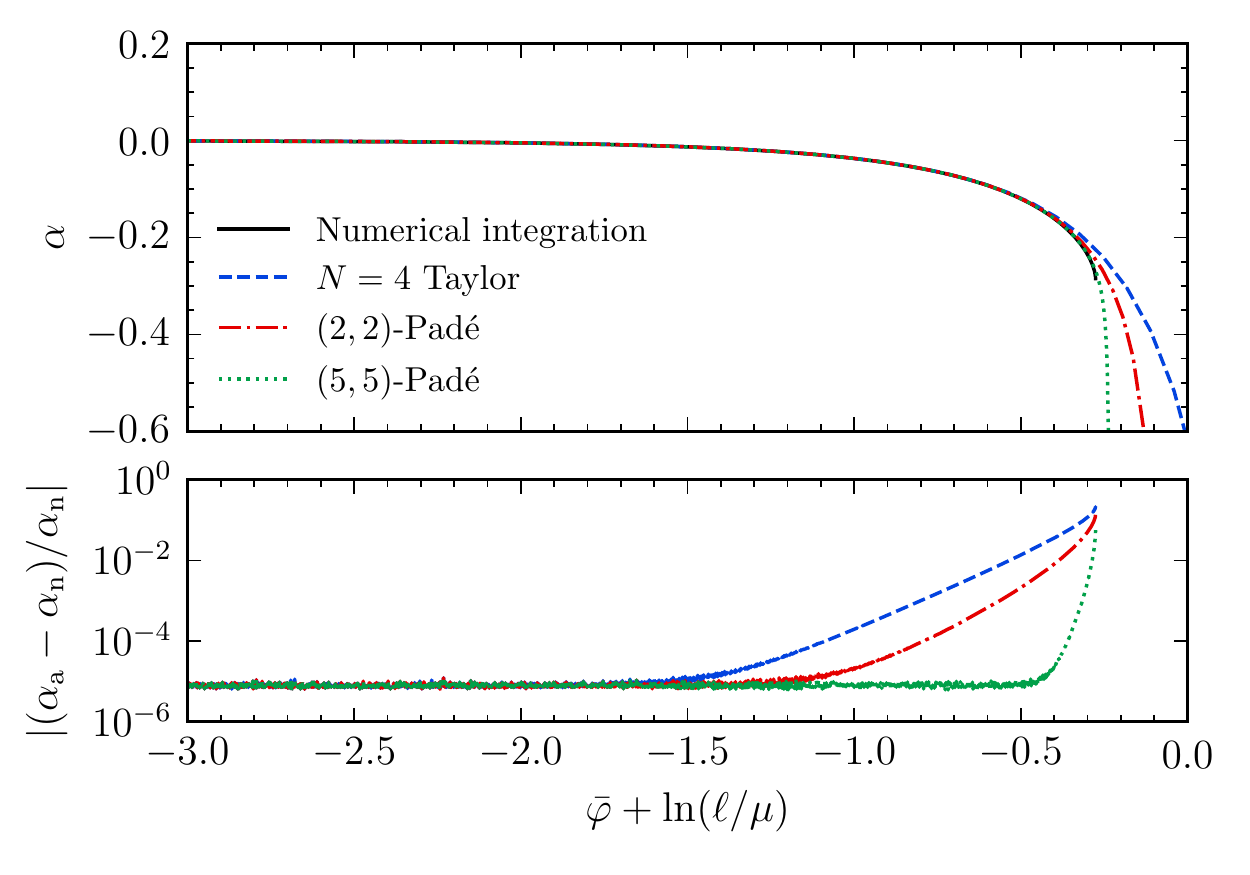}
\caption{
Black hole mass and sensitivity in the dilatonic theory \eqref{eq:def_dilatonic_theory} as
functions of the quantity $\vpbg + \ln(\ell / \mu)$ introduced in Eq.~\eqref{eq:combinationDilaton}.
Left panel: the numerical ADM-to-irreducible mass ratio $M/\mu$.
Right panel: the numerical sensitivity $\alpha$ and its analytic estimates from Eq.~\eqref{eq:analytical_sens} with $N=4$, its $(2,2)$-Pad\'e resummation, and the $(5,5)$-Pad\'e resummation of Eq.~\eqref{eq:analytical_sens}
with $N=10$.
The bottom-right panel shows the fractional error between analytic (``a'') and numerical (``n'') calculations.
The numerical sensitivity and its ($5,5$)-Pad\'e counterpart show excellent agreement, expect for one substantial qualitative difference: the Pad\'e approximants are singular as an artifact of the method, while the numerical sensitivity curve ends at $\vpbg + \ln(\ell / \mu)\approx -0.276$ as we approach saturation of Eq.~\eqref{eq:dl_bound}.
In the limit $\vpbg + \ln(\ell / \mu)\to -\infty$ we have $M/\mu\to1$ and $\alpha\to 0$, and at the end points we find $M/\mu\approx 0.913$ and $\alpha_{A} \approx - 0.285$.
}
\label{fig:dil_sensitivity_vs_phi0}
\end{figure*}

In Fig.~\ref{fig:dil_sensitivity_vs_phi0} we show the ADM-to-irreducible mass
ratio $M/\mu$ (left panel) and sensitivity $\alpha$ (right panel) as functions
of the combination~\eqref{eq:combinationDilaton}.
As with Fig.~\ref{fig:ss_sensitivity_vs_phi0}, the top-right panel also
includes analytic estimates of $\alpha$ derived from the Taylor series~\eqref{eq:analytical_sens} with $N=4$,
its $(2,2)$-Pad\'e resummation~\cite{Julie:2019sab}, and the $(5,5)$-Pad\'e resummation of
Eq.~\eqref{eq:analytical_sens} extended to $N=10$ here, with $x=\lsqr e^{2\vpbg}/(2\mu^2)$.
We discuss the sensitivity $\beta$, obtained from $\alpha$ through
Eq.~\eqref{eq:sensitivity_beta}, in Appendix~\ref{app:beta_sensitivity}.

As shown by the bottom panel, the agreement between
the numerical sensitivity and its $(5,5)$-Pad\'e counterpart is excellent, except for one substantial qualitative difference.
The Pad\'e approximants feature artificial poles, while the numerical sensitivity is finite and its curve terminates earlier than that of the $(5,5)$-Pad\'e approximant, at
\begin{equation}
\vpbg +  \ln\left( \frac{\ell}{\mu} \right)\lesssim-0.276\,.\label{eq:dil_max_scalar_Infty}
\end{equation}
We find that the saturation of this inequality indeed coincides numerically with that of the horizon bound \eqref{eq:dl_bound}, which, in turn, indicates that a hidden curvature singularity is approaching the black hole's horizon, in analogy with the shift-symmetric theory.

The role of the scalar background $\vpbg$ of a fixed Wald entropy black hole
therefore resembles the shift-symmetric case:
\begin{enumerate}
\item
when $\vpbg\to -\infty$, the black hole reduces to the Schwarzschild solution, since
$M/\mu\to 1$ and $\alpha=-Q/M\to 0$ (as well as its derivatives $\beta$), both analytically and numerically;
\item
when $\vpbg$ is increased, the sensitivity is negative and a hidden curvature
singularity approaches the horizon at $\vpbg + \ln(\ell / \mu) \approx -0.276$,
with $M/\mu\approx 0.913$ and $\alpha\approx -0.285$, as shown in
Fig.~\ref{fig:dil_sensitivity_vs_phi0}: cf.~\eqref{eq:dil_max_scalar_Infty}.
\label{itm:dil_ext}
\end{enumerate}
The impact of point~\ref{itm:dil_ext}~above on adiabatically inspiralling black
hole binaries will be studied in Sec.~\ref{sec:application}.

\subsection{Gaussian theory}
\label{sec:coup_gauss}

As a third and last example, consider the theory introduced in Ref.~\cite{Doneva:2017bvd},
\begin{equation}
    f(\vp) = - \frac{1}{12} \exp(- 6 \vp^2)\,,
    \label{eq:coupling_gauss}
\end{equation}
for which the action~\eqref{eq:action} is invariant under the $\mathds{Z}_{2}$-symmetry transformation $\vp \to -\vp$.
We note that the Schwarzschild spacetime is a solution of this theory when $\vp = 0$, since then Eq.~\eqref{eq:eom_metric} reduces to $R_{\mu\nu}=0$ and $f_{,\vp}(\vp)$ vanishes in Eq.~\eqref{eq:eom_scalar}.

For small $\vp$, the coupling can be approximated as
\begin{align}
 f(\vp)=\frac{1}{2} \vp^2+\dots
\label{eq:coupling_quad}
\end{align}
modulo boundary terms in the action,
which is the quadratic model studied in Ref.~\cite{Silva:2017uqg} and
developed further in Refs.~\cite{Silva:2018qhn,Macedo:2019sem}.
When the ratio
\begin{equation}
    \ell / M \approx 1.704\,, \quad \textrm{or} \quad
    \lstar=\ell / r_H \approx 0.852
    \label{eq:scalarization_threshold}
\end{equation}
is exceeded, the Schwarzschild spacetime is unstable, and ``spontaneously scalarized'' black
holes with nontrivial scalar field profiles branch off from
the Schwarzschild solution~\cite{Doneva:2017bvd,Silva:2017uqg}.
In the full Gaussian theory \eqref{eq:coupling_gauss}, scalarized yet stable
black holes were obtained numerically \cite{Doneva:2017bvd}, but they are
restricted to asymptotically zero scalar fields.
In Appendix~\ref{app:find_scalarized} we review how these scalarized solutions were found.

Here, we derive the numerical sensitivity of a scalarized black hole as a function of $\ell/\mu$ and of its generically \textit{nonzero} asymptotic scalar field value $\vpbg$.
Note that a scalarized black hole's sensitivity cannot be estimated from the analytic formula \eqref{eq:analytical_sens}, which can only vanish when $\vpbg=0$, since then $f'(\vpbg)=0$.

The condition for the existence of a real scalar field at the horizon of constant entropy black holes \eqref{eq:scalarh_bound_sw} is
\begin{equation}
\frac{\lsqr e^{-6\varphi_H^2}}{2\mu^2}\left(\sqrt{6}|\varphi_H|-\frac{1}{12}\right) <1\,,
\label{eq:reg_gauss_rH}
\end{equation}
which is a transcendental equation for $\vp_H$, while the Wald entropy \eqref{eq:wald} reads
\begin{equation}
    \sS_{\rm W} = \pi [r_{H}^2 - (\lsqr\! / 3) \exp(-6 \vp_{H}^{2})]\,.
    \label{eq:wald_gauss}
\end{equation}
Using the $\mathds{Z}_{2}$ symmetry of the theory, we anticipate from the definition \eqref{eq:sensitivity} that $\alpha\to-\alpha$ when $\vpbg\to-\vpbg$.
However, the Gaussian theory lacks
further symmetries to obtain $\alpha$ at once for all $\ell/\mu$ ratios, contrary to the shift-symmetric and dilatonic theories.
Thus, we focus here on a few illustrative examples, but gather our complete results for values
$(\ell/\mu)^2\leqslant 20$ with increment $\Delta(\ell/\mu)^2 \approx 0.2$
in~\cite{FLJRepo}.
We leave a discussion of the sensitivity $\beta$ to Appendix~\ref{app:beta_sensitivity}.

\begin{figure*}[ht]
\includegraphics[width=\columnwidth]{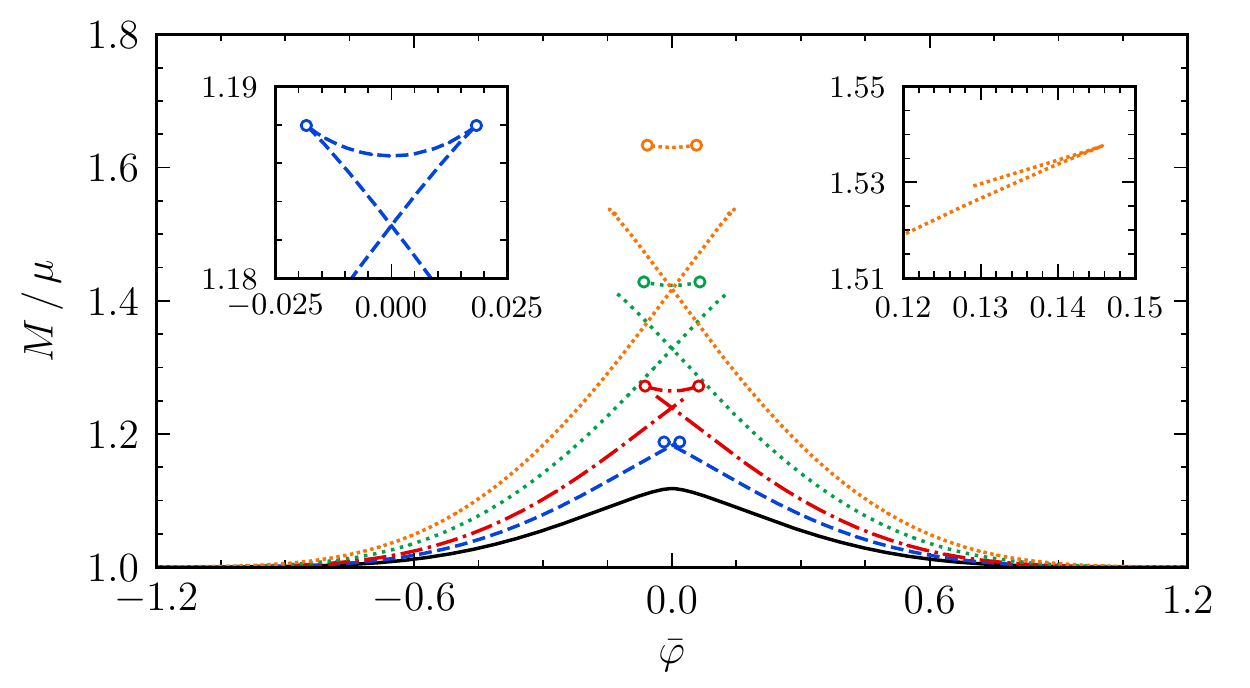}
\includegraphics[width=\columnwidth]{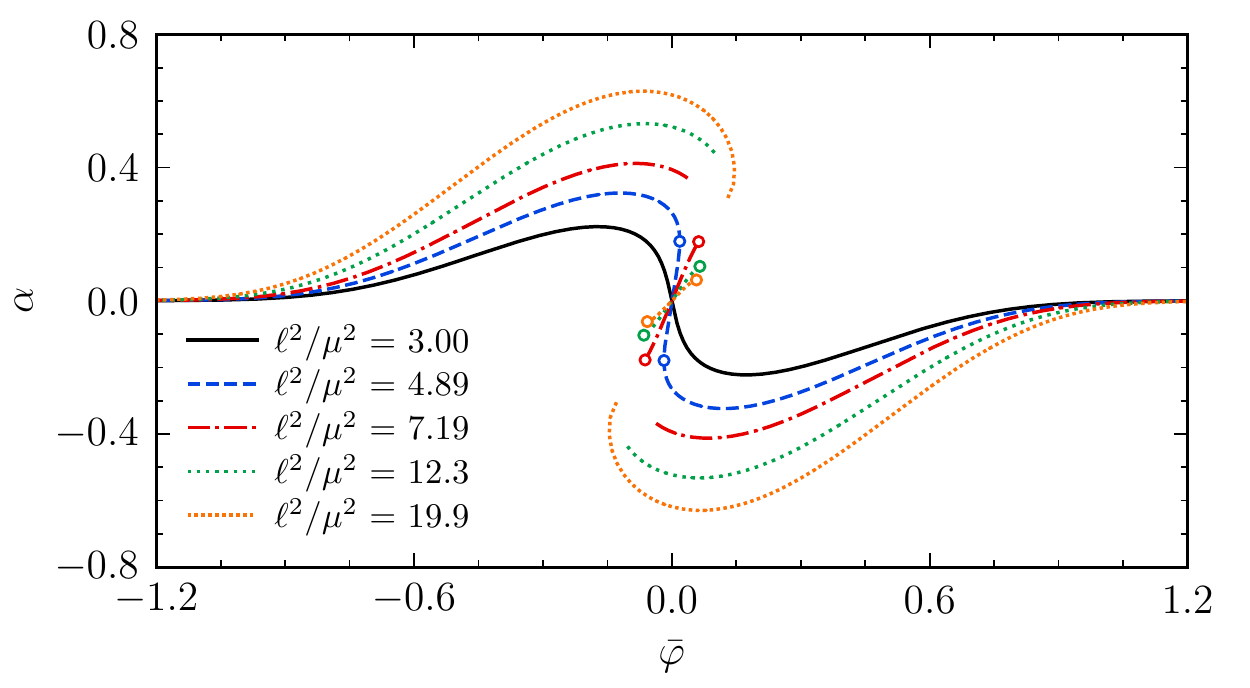}
\caption{
Black hole mass and sensitivity in the Gaussian theory \eqref{eq:coupling_gauss} as
functions of $\vpbg$ and $\ell/\mu$. We consider five families of constant entropy
solutions with $(\ell / \mu)^{2} = \{ 3.00,\, 4.89,\, 7.19,\, 12.3,\, 19.9\}$, and the
legend is shared by both panels.
Left panel: the numerical ADM-to-irreducible mass ratio $M/\mu$.
Right panel: the numerical sensitivity $\alpha$.
For values $(\ell/\mu)^2\lesssim 3.82$ the curves include a stable Schwarzschild solution $\alpha=0$ at $\vpbg=0$.
For values $(\ell/\mu)^2 \gtrsim 3.82$ the curves
are multivalued due to the occurrence of spontaneous scalarization.
The segments bracketed by the circles denote the ``Schwarzschild branches'', which
include an unstable Schwarzschild solution $\alpha=0$ at $\vpbg=0$.
The remainder of the curves form two ``scalarized branches'' which include the
stable scalarized black holes presented in Appendix~\ref{app:find_scalarized} and reviewed in Table~\ref{tab:scalarized_bhs_props}, with $\vpbg = 0$ but $\alpha\neq 0$ .
For $(\ell/\mu)^2\in[3.82,5.61]$, these three branches are connected.
At their junction, $M/\mu$ features a cusp, and the slope of $\alpha$ is infinite, cf. $(\ell/\mu)^2=4.89$ and the left inset;
but when $(\ell/\mu)^2\gtrsim 5.61$, the branches are discontinuous due to the violation of the horizon bound~\eqref{eq:reg_gauss_rH}.
When $(\ell/\mu)^2\in[13.6,13.8]$ and $(\ell/\mu)^2\gtrsim 14.0$, the ratio $M/\mu$ of scalarized branches also features a cusp, cf. $(\ell/\mu)^2=19.9$ and the right inset.
For all $\vpbg$ values, $M / \mu$ is larger along the Schwarzschild branch than along its scalarized counterparts, and at $\vpbg=0$, both scalarized branches have the same $M / \mu$ ratio.
}
\label{fig:gaussian_l6_sens_trans}
\end{figure*}

In Fig.~\ref{fig:gaussian_l6_sens_trans}, we show the ADM-to-irreducible mass ratio $M/\mu$ (left panel) and sensitivity
$\alpha$ (right panel) for five values of the ratio $\ell/\mu$.
For sufficiently small $\ell / \mu$ the curves are single-valued, as shown
by the example $(\ell / \mu)^2 = 3.00$.
The sensitivity is qualitatively similar to the analytic curves obtained in Ref.~\cite{Julie:2019sab}, Fig.~3, since spontaneous scalarization does not occur at $\vpbg=0$ in such
cases.
Moreover, we always find that $\alpha=0$ at $\vpbg=0$, thus recovering a stable Schwarzschild spacetime $\varphi=0$ at this particular point: see below Eq.~\eqref{eq:coupling_gauss}.

However, when
\begin{equation}
(\ell/\mu)^2 \gtrsim 3.82\,,
\end{equation}
the scalarization threshold \eqref{eq:scalarization_threshold} is exceeded [take $\mu^2=\sS_{\rm W}/4\pi$, where $\sS_{\rm W}$ is given by Eq.~\eqref{eq:wald_gauss} and $\varphi_H=0$ for Schwarzschild], and the situation changes.
First, the curves become \emph{multivalued}.
This is most easily seen for $\vpbg = 0$, such that $\alpha$ can
either be zero, or take two equal and opposite nonzero values, whose
magnitude increase with $\ell / \mu$.
The former vanishing $\alpha$ corresponds to an unstable Schwarzschild spacetime, while its latter nonzero values are those of the stable scalarized black holes reviewed in Appendix~\ref{app:find_scalarized}, with asymptotically zero scalar fields $\vpbg=0$.
For clarity, we gather in Table~\ref{tab:scalarized_bhs_props} the correspondence between the values $\ell/\mu$ used here and those of the ratio $\lstar=\ell/r_H$ used in the literature and in Appendix~\ref{app:find_scalarized}, found using Eq.~\eqref{eq:a_mu_ratio}.

For larger $\ell / \mu$ ratios, the sensitivity curve is increasingly sheared
and it can even be discontinuous when $(\ell/\mu)^2\gtrsim 5.61$, see e.g. $(\ell /
\mu)^2 = 7.19$.
The discontinuity happens due to the existence of intervals of values of $\vp_H$ which
do not satisfy the horizon bound~\eqref{eq:reg_gauss_rH}, but that are encountered while
implementing the algorithm given in the beginning of Sec.~\ref{sec:bh_sens}.
These intervals are shown in the bottom panel of Fig.~\ref{fig:find_scal_sols}, in
Appendix~\ref{app:find_scalarized}.
As one approaches the saturation of Eq.~\eqref{eq:reg_gauss_rH}, a hidden curvature
singularity resembling that of shift-symmetric theories (cf.
Fig.~\ref{fig:ss_sensitivity_vs_phi0_bh_props}) approaches the horizon.
Finally, as $\vpbg \to \pm \infty$, we notice that $M/\mu\to 1$ and $\alpha\to 0$ (as well as $\beta$ and higher derivatives) for all $\ell/\mu$ values, thus recovering scalar-field-decoupled black holes.

Given a fixed ratio $(\ell/\mu)^2 \gtrsim 3.82$, we will split our curves into the three following segments, or ``branches''.
We name the branch bracketed by the circles in
Fig.~\ref{fig:gaussian_l6_sens_trans}, going through $\alpha=0$ at $\vpbg=0$
and with the largest $M/\mu$ ratio at $\vpbg=0$, the ``Schwarzschild branch''.
It describes a family of black holes that can be continuously deformed into
the Schwarzschild solution through adiabatic changes in $\vpbg$.
We recall that since $(\ell/\mu)^2 \gtrsim 3.82$, the Schwarzschild
spacetime (with $\varphi=0$) is unstable~\cite{Doneva:2017bvd,Silva:2017uqg}.
However, the other points with $\vpbg\neq 0$ belonging to this branch correspond to
new black hole spacetimes whose stability is so far unknown. We leave their
study to future work.
Next, we name the remaining two branches, going through equal and opposite $\alpha\neq 0$ and equal $M/\mu$ at $\vpbg=0$, the ``scalarized branches''. They describe two families of black holes that can be continuously deformed into each other through adiabatic changes of $\vpbg$, and that include, e.g., the stable scalarized black holes listed in Table~\ref{tab:scalarized_bhs_props} at $\vpbg=0$.
The points with $\vpbg\neq 0$ belonging to these branches also represent new
black hole solutions, whose stability we also leave to future work.
\begin{table}[t]
\begin{tabular}{ c  c  c  c}
\hline
\hline
$\lstar^2=\lsqr / r_H^2$ & $\lsqr / \mu^{2}$ & $\pm \, \vp_H$ & $\pm \, \alpha$   \\
\hline
1.00  &  4.89  &  0.318  &  0.264  \\
1.56  &  7.19  &  0.481  &  0.394  \\
2.78  &  12.3  &  0.614  &  0.522  \\
4.58  &  19.9  &  0.702  &  0.618  \\
\hline
\hline
\end{tabular}
\caption{
Scalarized black hole examples in the theory \eqref{eq:coupling_gauss}, with asymptotically vanishing scalar fields, $\vpbg=0$.
The values in the first three columns are related to each other by Eq.~\eqref{eq:a_mu_ratio}.
}
\label{tab:scalarized_bhs_props}
\end{table}

For $(\ell/\mu)^2\in[3.82, 5.61]$, the three branches above are
connected, see the $(\ell/\mu)^2=4.89$ curves in
Fig.~\ref{fig:gaussian_l6_sens_trans}.
Hence, in principle, black holes can evolve adiabatically from one
branch to another.
Note however that $M/\mu$ features a cusp at the branches' junction, see the left inset in Fig.~\ref{fig:gaussian_l6_sens_trans}. The black hole's
sensitivity $\beta$, defined as the slope of $\alpha$ by Eq.~\eqref{eq:sensitivity_beta}, must
therefore diverge at the junction, as shown in Appendix~\ref{app:beta_sensitivity}.
Finally, for values $(\ell/\mu)^2\in[13.6, 13.8]$ and
$(\ell/\mu)^2\gtrsim 14.0$, the ratio $M/\mu$ of the scalarized branches also features a
cusp (while the three branches are always disconnected), as shown by the example $(\ell/\mu)^2=19.9$ in the right inset of Fig.~\ref{fig:gaussian_l6_sens_trans}.

Let us conclude this section with the following observation, which will play an important role below.
Consider a scalarized black hole with fixed $(\ell/\mu)^2 \gtrsim 3.82$ and,
initially, $\alpha>0$ ($\alpha<0$) at $\vpbg=0$.
Then, $\vpbg$ cannot be increased (decreased) in adiabatic conditions
indefinitely.
Indeed, depending on $(\ell/\mu)^2$: either
\begin{enumerate}
\item
the black hole flows along the scalarized branch up to a cusp of $M/\mu$, cf. $(\ell/\mu)^2=4.89$ or $(\ell/\mu)^2=19.9$ in Fig.~\ref{fig:gaussian_l6_sens_trans}. At the cusp, $\vpbg$ cannot be increased (decreased) further, or the black hole must leave its branch discontinuously, thus losing adiabaticity; or
\label{itm:gauss_adiab}
\item
the black hole eventually reaches the end point of its scalarized branch,  cf. $(\ell / \mu)^2 = 7.19$ and $(\ell / \mu)^2 =12.3$ in Fig.~\ref{fig:gaussian_l6_sens_trans}. At the end point, the condition \eqref{eq:reg_gauss_rH} is saturated and a hidden singularity approaches the black hole's horizon.
\label{itm:gauss_ext}
\end{enumerate}
The consequences of points~\ref{itm:gauss_adiab}~and~\ref{itm:gauss_ext}~above on adiabatically inspiralling black hole binaries
will be studied in Sec.~\ref{sec:application}.

\begin{figure}[t]
\includegraphics[width=.85\columnwidth]{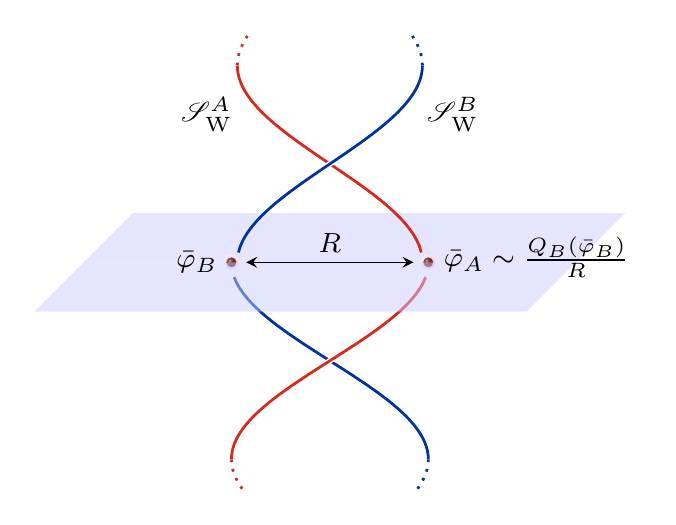}
\caption{Illustration of the two-body dynamics.
Two black holes with ADM masses and scalar charges $M_{A}(\vpbg_{A})$,  $Q_{A}(\vpbg_{A}) = - M_{A}(\vpbg_{A}) \alpha_{A}(\vpbg_{A})$ and their $B$ counterparts
are inspiralling at fixed irreducible masses $\mu_A$ and $\mu_B$.
In the previous sections, we calculated these quantities non perturbatively.
The background scalar field $\vpbg_{A}$, experienced by black hole $A$,
can now be estimated at leading-order by the $\propto 1/R$
potential~\eqref{eq:phi0PN_A} sourced by black hole $B$, and vice-versa.
The values of $\vpbg_{A}$ and $\vpbg_{B}$ change as the orbital radius $R$ decreases.
}
\label{fig:bbh_diagram_more_tech}
\end{figure}

\section{The fate of black hole binaries}
\label{sec:application}

Perhaps the most startling conclusion we drew from Figs.~\ref{fig:ss_sensitivity_vs_phi0} to~\ref{fig:gaussian_l6_sens_trans} above is that adiabatic changes to the
environmental scalar field $\vpbg$ of a black hole can
induce it to evolve towards a limiting $\vpbg$ value,
beyond which it can no longer be continuously deformed into a black hole with the same Wald entropy.

We can then ask: could this scenario be realized in a black hole binary, where changes to the scalar environment $\vpbg_A$ of a black hole $A$ are induced by the scalar hair of an inspiralling companion $B$? The setup is illustrated in Fig.~\ref{fig:bbh_diagram_more_tech}.

To answer this question, we use the results of Ref.~\cite{Julie:2019sab}.
There, the PN dynamics of bound binary systems was studied in the weak-field,
small orbital velocity limit $\mathcal O(M/R)\sim\mathcal O(v^2)$.
The field equations were solved iteratively around a Minkowski metric
$g_{\mu\nu}=\eta_{\mu\nu}+\delta g_{\mu\nu}$, and a constant scalar background $\varphi_0$ imposed by the binary's cosmological
environment, $\varphi=\varphi_0+\delta\varphi$.
At Newtonian (0PN) order, to which we restrict ourselves here, we thus have~\cite{Julie:2019sab}
\begin{subequations}
\label{eq:phi0PN}
\begin{align}
\vpbg_A &= \varphi(t,\bm{x}_A)=\varphi_0-\frac{M_B^0\alpha_B^0}{R}+\mathcal O(v^4)\,,\label{eq:phi0PN_A}\\
\vpbg_B &= \varphi(t,\bm{x}_B)=\varphi_0-\frac{M_A^0\alpha_A^0}{R}+\mathcal O(v^4)\,,
\end{align}
\end{subequations}
where $R=|\bm{x}_A-\bm{x}_B|$ is the orbital separation, and where the
superscript ``$0$'' denotes a quantity evaluated by formally setting $\vpbg_{A,B}=\varphi_0$.
For shift-symmetric and dilatonic models, we can set $\varphi_0=0$ without loss of generality,
using the symmetries given below Eq.~(\ref{eq:def_shift_sym_theory}) and below
Eq.~\eqref{eq:def_dilatonic_theory}.
For the Gaussian theory we choose $\varphi_0=0$, which corresponds to a
nondynamical scalar field on cosmological scales, at least
classically~\cite{Anson:2019uto}.
Given a binary system with irreducible masses $\mu_A$ and $\mu_B$, and a fundamental coupling value $\ell$, the quantities
entering Eqs.~(\ref{eq:phi0PN}) are then fully evaluated from
Figs.~\ref{fig:ss_sensitivity_vs_phi0},~\ref{fig:dil_sensitivity_vs_phi0} and~\ref{fig:gaussian_l6_sens_trans}
by setting formally $\vpbg=0$ there, and they hence depend
only on the ratios $\ell/\mu_A$ and $\ell/\mu_B$.

\begin{figure*}[t!]
\includegraphics[width=0.85\columnwidth]{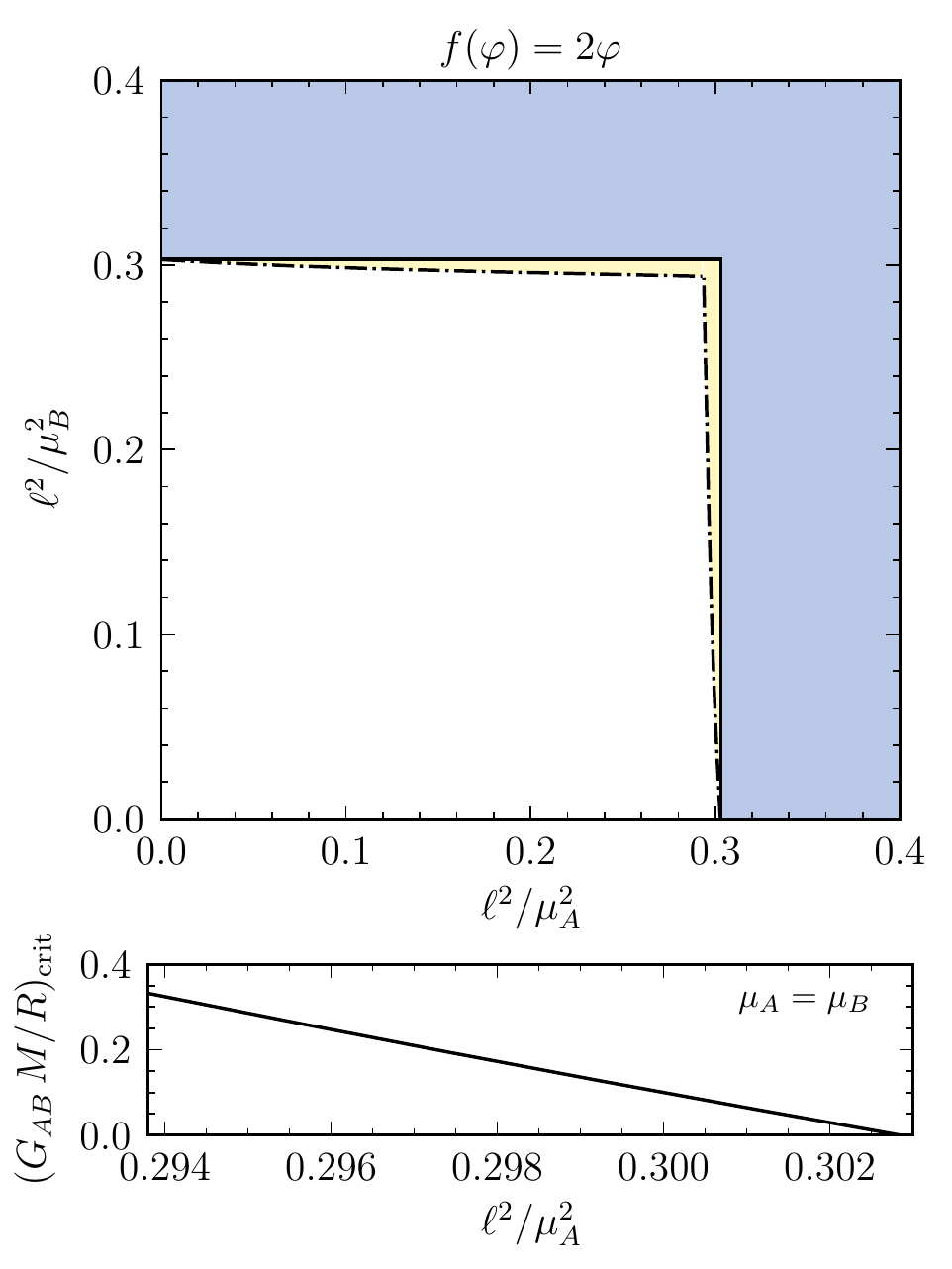}
\hspace{0.75cm}
\includegraphics[width=0.85\columnwidth]{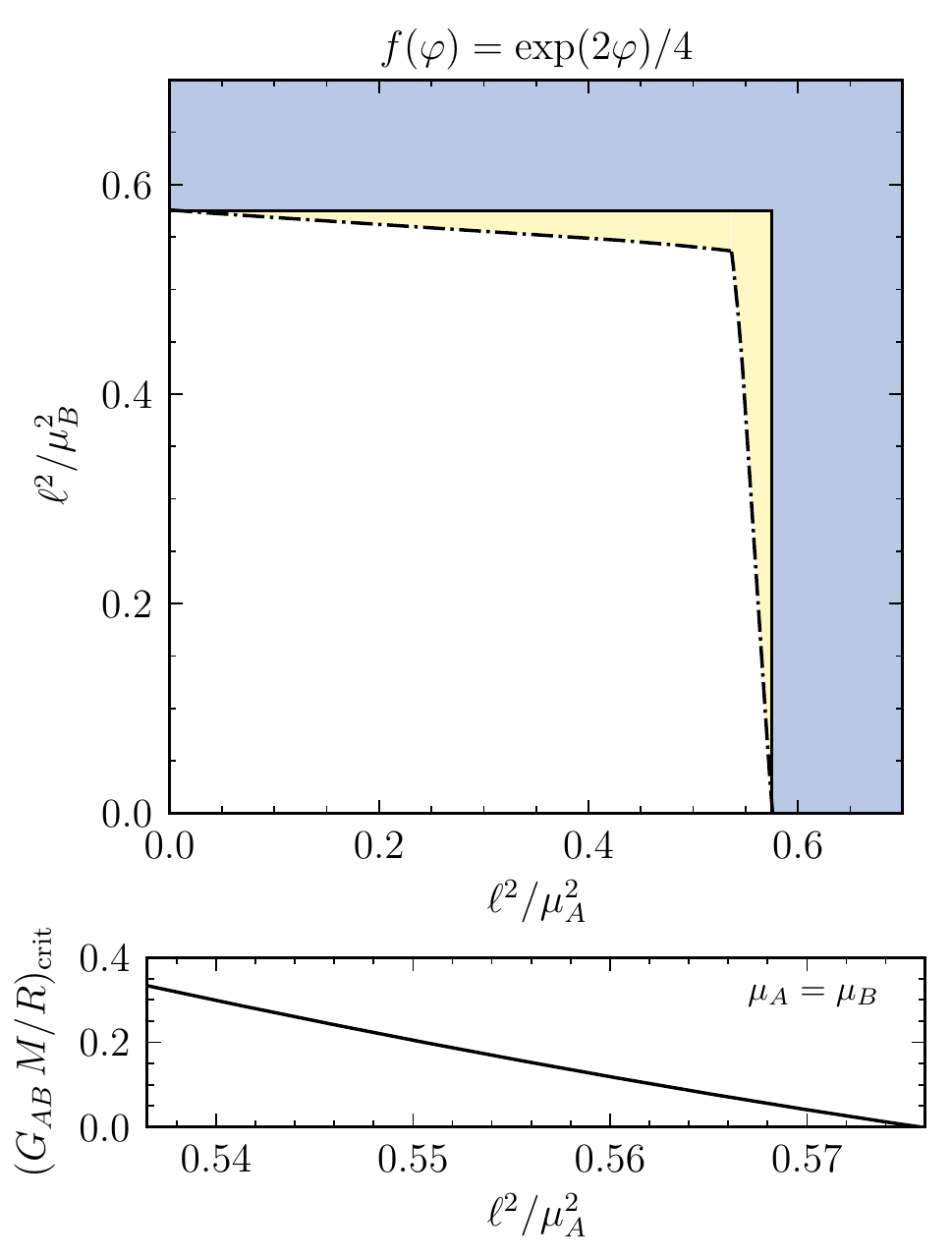}
\caption{
Parameter space of an inspiralling binary black hole with fixed ratios $\ell/\mu_A$ and $\ell/\mu_B$ in shift-symmetric and dilatonic theories.
In the infinite separation limit $R\to\infty$, each black hole is isolated and
the conditions \eqref{eq:condition_inspiral_shift_sym} and
\eqref{eq:condition_inspiral_dilaton} exclude the upper panels' dark (blue) shaded
regions.
In the light (yellow) shaded regions,
the binary is initially
regular at infinity, but at least one of the black holes violates the conditions above
before the system reaches its light ring \eqref{eq:light_ring}, i.e., at some
critical orbital radius $R_{\rm crit}>R_{\rm LR}$.
This signals that this black hole's hidden curvature singularity has reached its horizon.
The critical orbital radius $R_{\rm crit}$ is shown by the
lower panels for the example of symmetric binary systems $\mu_A=\mu_B$, and it varies between $R_{\rm crit}\to \infty$ and $R_{\rm crit}=R_{\rm LR}$, depending on how close the black holes initially are to saturating \eqref{eq:condition_inspiral_shift_sym} or
\eqref{eq:condition_inspiral_dilaton}.
Left panels: in the shift-symmetric case $f(\varphi)=2\varphi$, isolated black
holes must satisfy $(\ell/\mu_A)^2\lesssim0.303$. For binaries, this bound is
tightened and becomes $(\ell/\mu_A)^2\lesssim0.294$ when $\mu_A=\mu_B$.
Right panels: in the dilatonic case $f(\varphi)=\exp(2\varphi)/4$, isolated
black holes must satisfy $(\ell/\mu_{A})^2\lesssim0.576$. For binaries, this bound is
tightened and becomes $(\ell/\mu_A)^2\lesssim0.536$ when $\mu_A=\mu_B$.
}
\label{fig:inspiral_constraints}
\end{figure*}

As a minimal value for the orbital radius, we use the light ring $R_{\rm LR}$.
Indeed, in general relativity the light ring marks the transition to the ringdown phase in a
compact binary evolution, and it can be estimated, e.g., using the
effective-one-body (EOB) formalism~\cite{Buonanno:1998gg,Buonanno:2000ef}.
The EOB framework was generalized to scalar-tensor theories in
Refs.~\cite{Julie:2017pkb,Julie:2017ucp}, but the results were shown in
Ref.~\cite{Julie:2019sab} to also include ESGB models.
Here we will need the light ring at 0PN,
\begin{equation}
R_{\rm LR}=3\,G_{AB}M\,,\label{eq:light_ring}
\end{equation}
where $M=M_A^0+M_B^0$, and $G_{AB}=1+\alpha_A^0\alpha_B^0$ is an effective
gravitational coupling reflecting the linear addition of the metric and scalar
interactions at this order.

In the following, we explore whether adiabatically readjusting black holes can be driven outside their domain of existence in an inspiralling binary black hole system with orbital radius
$R>R_{\rm LR}$ for shift-symmetric, dilatonic and Gaussian models.

\subsection{Shift-symmetric theory}

When $f(\varphi)=2\varphi$,
we have from Fig.~\ref{fig:ss_sensitivity_vs_phi0}
that both black holes in a binary must satisfy the condition \eqref{eq:ss_max_scalar_Infty}:
 \begin{subequations}
\begin{align}
&\vpbg_A-\frac{\mu_A^2}{2\lsqr}\lesssim-1.651\,, \,\,\,
 \label{eq:condition_inspiral_shift_symA}\\
&\vpbg_B-\frac{\mu_B^2}{2\lsqr}\lesssim-1.651\,.
\end{align}\label{eq:condition_inspiral_shift_sym}%
 \end{subequations}

In the early inspiral regime, $R\to\infty$ and both $\vpbg_A$ and
$\vpbg_B$ vanish, see Eqs.~\eqref{eq:phi0PN} and below.
The conditions above then yield $(\ell/\mu_{A})^2\lesssim0.303$ and
$(\ell/\mu_{B})^2\lesssim0.303$, which excludes the dark (blue) shaded region
in the upper left panel of Fig.~\ref{fig:inspiral_constraints}.
In this regime, the Gauss-Bonnet coupling $\ell$ is bounded from above: it must be
smaller than a fraction of each black holes' fixed Wald entropies
$\sS_{\rm W}^A=4\pi\mu_A^2$
and
$\sS_{\rm W}^B=4\pi\mu_B^2$.
Our result is consistent with previous constraints obtained in, e.g.,
Ref.~\cite{Witek:2018dmd} for isolated black holes with constant ADM masses.

However, in general $\vpbg_A$ and $\vpbg_B$ are nonzero and positive
[cf.~Eq.~\eqref{eq:phi0PN}], $\alpha_A^0$ and $\alpha_B^0$ being negative and given in
Fig.~\ref{fig:ss_sensitivity_vs_phi0}, and they increase as the orbital radius $R$ decreases.
This effectively tightens the
conditions~\eqref{eq:condition_inspiral_shift_sym} gradually along the inspiral, and extends the excluded
parameter space, as depicted by the light (yellow) shaded region.
For any point in the latter, there indeed exists a critical orbital radius $R_{\rm crit}>R_{\rm LR}$
where at least one of the conditions~\eqref{eq:condition_inspiral_shift_sym}
is saturated. As discussed in Sec.~\ref{sec:coup_ssym}, this signals that at least one of the black holes' hidden singularity is approaching its horizon, see point \ref{itm:ss_ext} there. At the border with the white region, we have $R_{\rm crit}=R_{\rm LR}$.

The light (yellow) shaded region is here relatively narrow, because the influence of black
hole A on $\vpbg_B$ is limited by the
assumption $R>R_{\rm LR}$, and the fact that
$\alpha_A^0 \approx -0.350$  at most, see Fig.~\ref{fig:ss_sensitivity_vs_phi0}.
The region also shrinks when, say, $(\ell/\mu_A)^2\ll 1$, because then black hole
A decouples from the scalar field, and it cannot affect $\vpbg_B$ since $\alpha_A^0\to 0$.
Conversely, the region is thickest for symmetric binaries $\mu_A=\mu_B$,
yielding a tighter bound $(\ell/\mu_{A})^2\lesssim0.294$.
Finally, conditions~\eqref{eq:condition_inspiral_shift_sym} are always satisfied down to $R_{\rm LR}$
in the white region.

The lower left panel of Fig.~\ref{fig:inspiral_constraints} shows the 0PN
potential value $G_{AB}M/R_{\rm crit}$ at criticality, that is when
Eqs.~\eqref{eq:condition_inspiral_shift_sym} saturate, for the example of
symmetric binaries, $\mu_A=\mu_B$.
This potential is related to the binary's orbital velocity $\dot\phi=\dd \phi / \dd t$ via
Kepler's law (see e.g. Ref.~\cite{Julie:2018lfp}):
\begin{equation}
\frac{G_{AB}M}{R_{\rm crit}}=(G_{AB}M\dot\phi_{\rm crit})^{2/3}+\mathcal O(v^4)\,,
\end{equation}
and its numerical value varies from $1/3$ ($R_{\rm crit}=R_{\rm LR}$) to zero
($R_{\rm crit}\to\infty$) as we move along the line $\mu_A=\mu_B$, across the
light (yellow) shaded region of the upper panel.

\subsection{Dilatonic theory}

The steps presented above are now easily adapted to the dilatonic case. When
$f(\varphi)=(1/4) \exp(2\vp)$, we have from Fig.~\ref{fig:dil_sensitivity_vs_phi0} that a binary black hole must satisfy two copies of the condition \eqref{eq:dil_max_scalar_Infty}:
\begin{subequations}
\begin{align}
&\vpbg_A+\ln\left(\frac{\ell}{\mu_A}\right)\lesssim-0.276\,,\label{eq:condition_inspiral_dilatonA}\\
&\vpbg_B+\ln\left(\frac{\ell}{\mu_B}\right)\lesssim-0.276\ .
\end{align}\label{eq:condition_inspiral_dilaton}
\end{subequations}

In the limit $R\to\infty$, both $\vpbg_A$ and $\vpbg_B$
vanish \eqref{eq:phi0PN}, and the conditions above yield $(\ell/\mu_{A})^2\lesssim0.576$ and
$(\ell/\mu_{B})^2\lesssim0.576$. The resulting excluded dark (blue) shaded region is
shown in the upper right panel of Fig.~\ref{fig:inspiral_constraints}.
When the orbital radius $R$ is finite, $\vpbg_A$ and $\vpbg_B$
are nonzero and the conditions \eqref{eq:condition_inspiral_dilaton} extend
the excluded parameter space, as shown by the light (yellow) shaded region.
Just as with the shift-symmetric case, for each point in this region there
exists a critical orbital radius $R_{\rm crit}>R_{\rm LR}$ at which at least one of the
conditions \eqref{eq:condition_inspiral_dilaton} is saturated.
As discussed in Sec.~\ref{sec:coup_dilaton}, the latter signals that one of the black holes' hidden singularities is approaching its horizon, see point \ref{itm:dil_ext} there.
The region is thickest when $\mu_A=\mu_B$, in which case the Gauss-Bonnet
coupling is bounded by $(\ell/\mu_{A})^2\lesssim0.536$.

The critical Newtonian potential
$G_{AB}M/R_{\rm crit}$ is shown in the lower panel for $\mu_A=\mu_B$, and it varies from $1/3$ when $R_{\rm crit}=R_{\rm LR}$, to zero
when $R_{\rm crit}\to\infty$.

\subsection{Gaussian theory}

\begin{figure}[h!]
\includegraphics[width=0.85\columnwidth]{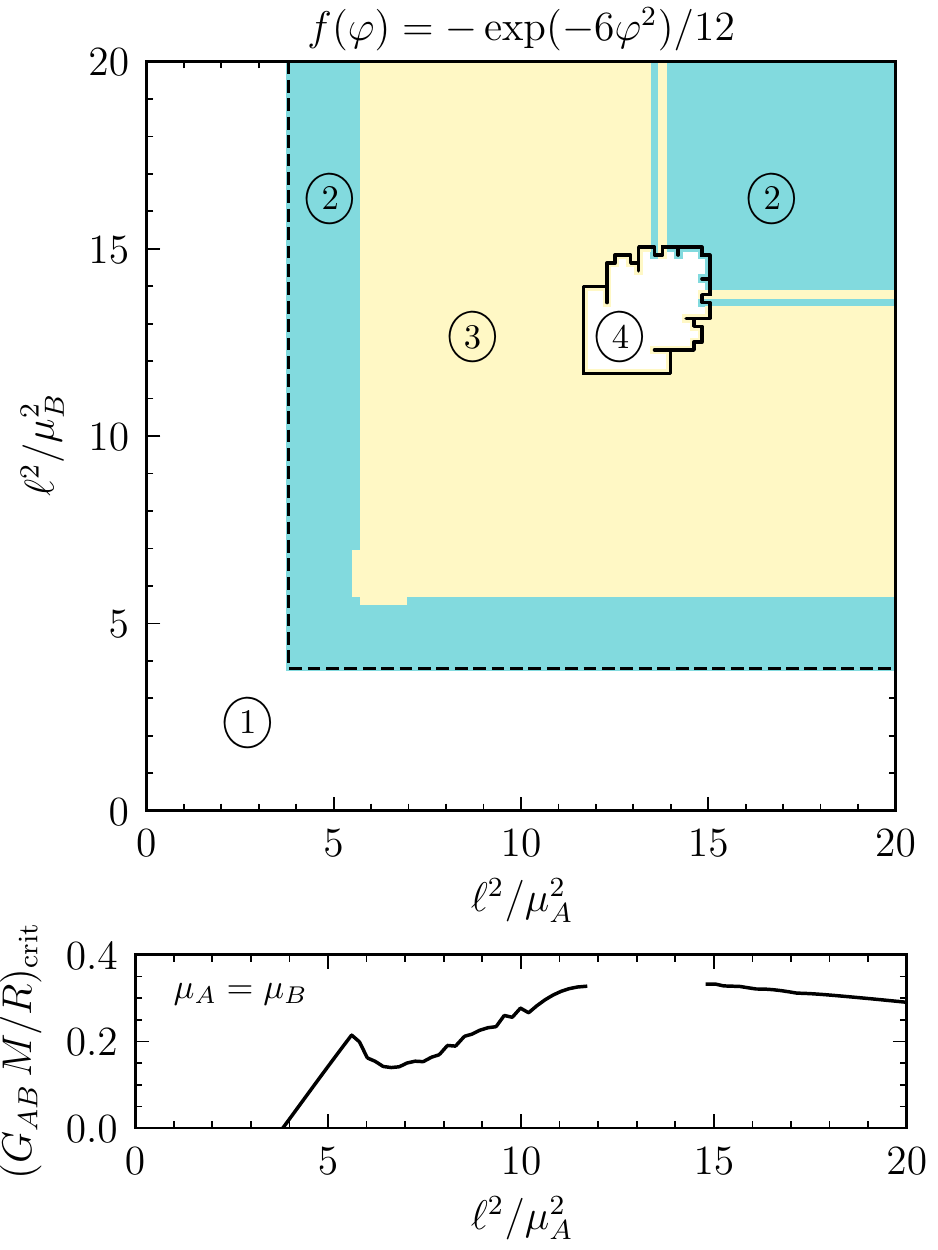}
\caption{
Parameter space of an inspiralling binary black hole with fixed $\ell/\mu_A$
and $\ell/\mu_B$ and initially vanishing scalar field environments
$\varphi_0=0$ in the Gaussian theory.
The black hole $A$ reduces initially to the Schwarzschild solution when
$(\ell/\mu_A)^2\lesssim 3.82$, and to a scalarized black hole when
$(\ell/\mu_A)^2\gtrsim3.82$.
In region \cn{1}, at least one black hole is initially Schwarzschild and the
system can inspiral adiabatically until merger.
When $(\ell/\mu_A)^2\gtrsim3.82$ and $(\ell/\mu_B)^2\gtrsim3.82$, each
inspiralling black hole evolves along a scalarized branch such as those
presented in Fig.~\ref{fig:gaussian_l6_sens_trans}.
When their sensitivities $\alpha_A$ and $\alpha_B$ have the same sign, they can
inspiral adiabatically until merger.
But when $\alpha_A$ and $\alpha_B$ have opposite signs, almost the entire
parameter space is excluded, as shown by the shaded regions.
In region \cn{2}, at least one of the black holes meets
point~\ref{itm:gauss_adiab} of Subsection~\ref{sec:coup_gauss} at some $R_{\rm crit}>R_{\rm LR}$
and must leave its scalarized branch discontinuously there.
In region \cn{3}, a black hole meets point~\ref{itm:gauss_ext} at $R_{\rm crit}>R_{\rm LR}$,
meaning that its hidden curvature singularity reaches its horizon there.
The remaining parameter space that allows the system to inspiral adiabatically
until merger is comparatively small, and depicted by region \cn{4}.
In the limit $\ell/\mu_A\gg\ell/\mu_B$, we have $|\alpha_A^0|\gg |\alpha_B^0|$ so that black hole $B$ always completes the scenario \ref{itm:gauss_adiab} or~\ref{itm:gauss_ext} before $A$. Then region \cn{2} is delimited by the ranges $(\ell/\mu_B)^2\in[3.82, 5.61]\,,[13.6, 13.8]$ and
$(\ell/\mu_B)^2\gtrsim 14.00$ discussed in
Subsection~\ref{sec:coup_gauss}.
The critical orbital radius $R_{\rm crit}>R_{\rm LR}$ at which points
\ref{itm:gauss_adiab} or \ref{itm:gauss_ext} indifferently happen is shown by
the lower panel for $\mu_A=\mu_B$ (but $\alpha_A=-\alpha_B$). It varies between
$R_{\rm crit}\to \infty$ and $R_{\rm crit}=R_{\rm LR}$.
}
\label{fig:inspiral_constraints_gaussian}
\end{figure}

The Gaussian case
is perhaps the most striking, but it must also be treated most carefully.
As discussed in Subsection~\ref{sec:coup_gauss}, when $(\ell/\mu_A)^2\gtrsim
3.82$ a black hole $A$ can in principle either belong to a Schwarzschild branch
or to one of its two scalarized counterparts.
However, in this paper we choose the quantity $\varphi_0$, which the environment
$\vpbg_A$ of the black hole reduces to when $R\to\infty$, to be zero [cf.
Eq.~\eqref{eq:phi0PN} and below]. Given such initial conditions, the black hole
must belong to a scalarized branch, since otherwise it would reduce initially
to an unstable Schwarzschild black hole.
When $(\ell/\mu_A)^2\gtrsim 3.82$, we therefore start from scalarized black
holes such as those presented in Table~\ref{tab:scalarized_bhs_props} and
typically discussed in the literature,
and when $(\ell/\mu_A)^2 \lesssim 3.82$, from stable Schwarzschild black holes.

Figure~\ref{fig:inspiral_constraints_gaussian} shows the parameter space of an
inspiralling black hole binary with fixed ratios $\ell^2/\mu_A^2$ and $\ell^2/\mu_B^2$,
which we explored for $(\ell/\mu_{A,B})^2\leqslant 20$ with increments in
$\Delta(\ell/\mu_{A,B})^2 \approx 0.2$.
In region \cn{1}, at least one of the black holes, say $A$, satisfies $(\ell/\mu_A)^2 \lesssim 3.82$. As shown by Fig.~\ref{fig:gaussian_l6_sens_trans} on the example $(\ell/\mu_A)^2=3.00$, we have $\alpha_A^0=0$ and thus $\vpbg_B=0$ by Eq.~\eqref{eq:phi0PN}. Since, moreover, black hole $A$ exists for all $\vpbg_A$ values, any point in region \cn{1} represents a binary system that can adiabatically inspiral until merger.

Next, we take $(\ell/\mu_A)^2\gtrsim 3.82$ and $(\ell/\mu_B)^2\gtrsim 3.82$, corresponding to two initially scalarized black holes such as those of Table~\ref{tab:scalarized_bhs_props}, which evolve along their respective scalarized branches as they inspiral, see Fig.~\ref{fig:gaussian_l6_sens_trans}.
We note that for every $\ell/\mu_{A,B}$ values we considered, the sensitivities at infinity $\alpha^0_{A,B}$ are always defined, contrary to shift-symmetric and dilatonic theories, which exclude the dark (blue) shaded regions of Fig.~\ref{fig:inspiral_constraints}.

Let us consider two scalarized black holes with sensitivities $\alpha_A$ and $\alpha_B$ of the same sign, taken to be positive without loss of generality.
Then, $\alpha_{A,B}^0>0$, so $\vpbg_{A,B}$ are both negative [cf.~Eq.~\eqref{eq:phi0PN}], with increasing magnitudes as the orbital radius $R$ decreases.
The black holes gradually drive each other away from the cusps or end points of their respective branches, see Fig.~\ref{fig:gaussian_l6_sens_trans}.
Hence, scalarized black hole binaries with sensitivities of the same sign can adiabatically inspiral until merger.

The picture above changes radically if the scalarized black holes $(\ell/\mu_A)^2\gtrsim 3.82$ and $(\ell/\mu_B)^2\gtrsim 3.82$ have sensitivities  $\alpha_A$ and $\alpha_B$ with opposite signs.
Indeed, if $\alpha_A^0>0$ and $\alpha_B^0<0$, then $\vpbg_A\geqslant0$ and $\vpbg_B\leqslant0$, with increasing absolute values as $R$ decreases.
We recover the situation described by points~\ref{itm:gauss_adiab} and~\ref{itm:gauss_ext} at the end of Sec.~\ref{sec:coup_gauss}.
As shown by the shaded regions of Fig.~\ref{fig:inspiral_constraints_gaussian}, the parameter space is then almost entirely excluded.
More precisely, for any point of region \cn{2}, there exists a critical orbital radius $R_{\rm crit}>R_{\rm LR}$ where at least one of the black holes, described by point~\ref{itm:gauss_adiab}, cannot inspiral further without leaving its branch discontinuously at $R_{\rm crit}$.
For any point of region \cn{3}, at least one black hole is described by point~\ref{itm:gauss_ext} and at some $R_{\rm crit}>R_{\rm LR}$, its hidden curvature singularity approaches the horizon.
In the limit $\ell/\mu_A\gg\ell/\mu_B$, $\alpha_A^0$ is large and we find that black hole $B$ always completes the scenario \ref{itm:gauss_adiab} or~\ref{itm:gauss_ext} before $A$.
The shaded regions are delimited
by $\ell/\mu_B$ intervals that reduce to those observed above points~\ref{itm:gauss_adiab} and~\ref{itm:gauss_ext} of Sec.~\ref{sec:coup_gauss}: when $(\ell/\mu_B)^2\in[3.82, 5.61]\,,[13.6, 13.8]$ and
$(\ell/\mu_B)^2\gtrsim 14.0$, the black hole $B$ is described by point~\ref{itm:gauss_adiab}, and in the complementary intervals it is described by point~\ref{itm:gauss_ext}.
The remaining allowed region \cn{4} is comparatively small. There, the black holes live on a sufficiently large scalar environment range, while keeping their scalar charges small enough for neither scenario~\ref{itm:gauss_adiab} nor~\ref{itm:gauss_ext} to happen.

The critical Newtonian potential $G_{AB}M/R_{\rm crit}$ at which
indifferently~\ref{itm:gauss_adiab} or~\ref{itm:gauss_ext} happen is
shown in the lower panel of Fig.~\ref{fig:inspiral_constraints_gaussian} for
$\mu_A=\mu_B$, and hence $\alpha_A=-\alpha_B$. It varies between $1/3$ when
$R_{\rm crit}=R_{\rm LR}$ and zero when $R_{\rm crit}\to\infty$.

\subsection{Epilogue}

In this section, we considered shift-symmetric, dilatonic and Gaussian ESGB
models. In all three cases, we found parameter spaces such that the adiabatic
inspiral of black hole binaries must break down. Let us conclude with the
following remarks.

First, we estimated the scalar environments $\vpbg_{A,B}$ of each black hole
using a Newtonian, leading-order approximation for simplicity.
Therefore, our results should not be considered as definitive, but they suggest
an interesting parameter space to be further explored, e.g., at higher PN
order~\cite{Julie:2019sab}, or using numerical
relativity~\cite{Witek:2018dmd,Okounkova:2020rqw,Silva:2020omi,East:2020hgw,East:2021bqk}
to reveal the ultimate fate of the black holes.
Note however that the phenomena we found can happen in the weak field regime.
As shown by the bottom panel of Fig.~\ref{fig:inspiral_constraints} in
shift-symmetric and dilatonic theories, a black hole $A$ can be adiabatically
driven to the end point of its sensitivity curve arbitrarily far into the
Newtonian regime $G_{AB}M/R\ll 1$, provided that $\ell/\mu_A$ is large enough.
In Gaussian theories, a black hole with ratio $\ell/\mu_A$ just above the scalarization threshold $(\ell/\mu_A)^2\approx 3.82$ must discontinuously leave its scalarized branch very early in the inspiral $G_{AB}M/R\ll 1$, see the bottom panel of Fig.~\ref{fig:inspiral_constraints_gaussian}.
As for the sensitivities $\alpha$, we recall that they were obtained nonperturbatively.

Second, the adiabatic analysis we performed describes binary systems in
the limit where tidal and out-of-equilibrium effects can be
discarded.
The fact that the adiabatic analysis formally breaks down might signal the occurrence of nonperturbative out-of-equilibrium phenomena.
It will hence be important to study the stability of the new black holes with
nonzero asymptotic scalar fields presented here.
In particular, addressing dynamical
(de)scalarization phenomena~\cite{Silva:2020omi} in ESGB gravity might
complete the scenario \ref{itm:gauss_adiab} found in
Subsection~\ref{sec:coup_gauss}.

Third, in all ESGB models considered, we found parameter space regions such that the hidden singularities of black holes can approach their horizons before merger, cf. point \ref{itm:gauss_ext} in Subsection~\ref{sec:coup_gauss} in the Gaussian case.
Unless the black holes then ``reopen'' into other compact objects~\cite{Kanti:2011yv}, the theories might simply not predict any binary evolution once~\ref{itm:gauss_ext} has happened.

If the predictions of this section are qualitatively confirmed in the future, while none of the scenarios listed above are observed in currently available and future gravitational wave event candidates, then new interesting constraints on ESGB theories might be obtained.
In particular, scalarized binary black holes with opposite scalar charges might be severely constrained.

\section{Conclusions}
\label{sec:conclusions}

We introduced a method to numerically calculate the
sensitivities of nonrotating black holes in ESGB theory.
This complements the analytical, but perturbative, calculation of
Ref.~\cite{Julie:2019sab}, which we also generalized here by calculating
higher-order terms in the perturbative series.
In the subclasses of this theory where comparison was possible, we showed that
analytical and numerical approaches agree remarkably well.
The numerical approach also allowed us to calculate the sensitivities of
spontaneously scalarized black holes for the first time.
We arrived, through a restrictive PN analysis, at the surprising conclusion
that adiabatically inspiralling black holes in some of these theories can in principle be driven outside
their domain of existence.
It would be interesting to confirm this finding by working to higher PN orders
or through numerical relativity
simulations~\cite{Witek:2018dmd,Okounkova:2020rqw,East:2020hgw,Silva:2020omi,East:2021bqk}.

Our results are important for the PN description of black hole binaries
in ESGB
gravity~\cite{Yagi:2011xp,Julie:2019sab,Shiralilou:2020gah,Shiralilou:2021mfl,Bernard:2018hta,Bernard:2018ivi},
including gravitational waveform predictions~\cite{Shiralilou:2021mfl,Bernard:2022noq}, allowing to finally specialize them to scalarized black hole binaries.
Our work could also be used to develop an effective action model~\cite{Khalil:2019wyy} of
dynamical \emph{black hole} descalarization~\cite{Silva:2020omi} and explore further
the differences with respect to \emph{neutron star} binaries in scalar-tensor
theories~\cite{Barausse:2012da,Palenzuela:2013hsa,Shibata:2013pra,Taniguchi:2014fqa,Sennett:2016rwa,Sennett:2017lcx} that predict
spontaneous scalarization~\cite{Damour:1993hw}.

More broadly, the method introduced here can, in principle, also be used to
calculate the sensitivities of black holes in other gravity theories, e.g., the
effective field theory introduced in~\cite{Cano:2019ore}, the effective
field theory for black hole scalarization of~\cite{Macedo:2019sem}, the
models of~\cite{Antoniou:2020nax,Ventagli:2020rnx,Antoniou:2021zoy}, and generalizations of
ESGB gravity with multiple scalar fields~\cite{Doneva:2020qww}.
Indeed, we expect the sensitivities, as
calculated here, to play a role
beyond ESGB theories: see Refs.~\cite{Cardenas:2017chu,Julie:2017rpw} for another example.
Hence, it is desirable that future work on black holes in
modified gravity theories study how the black hole ``charges''
vary as a function of the theory parameters, but also of the asymptotic value of the scalar field (if any) at fixed Wald entropy.

Our findings open some avenues for future work.
First, we could analyze the stability of the constant-entropy sequence of
solutions for the Gaussian theory studied in Sec.~\ref{sec:coup_gauss}.
It is known that the equations describing gravitational perturbations of such black holes
can cease to be hyperbolic~\cite{Blazquez-Salcedo:2018jnn,Blazquez-Salcedo:2020rhf,Blazquez-Salcedo:2020caw},
suggesting that their time evolution becomes ill-posed.
Taking this fact in consideration could in principle shrink further the
exclusion regions in Fig.~\ref{fig:inspiral_constraints_gaussian}, but more work is needed to draw definite
 conclusions.

Finally, in preparation to model the binary dynamics of spinning black holes in
ESGB gravity, one could extend the calculation done here to
rotating black holes.
The inclusion of spin would introduce a ``moment of inertia sensitivity''
analogous to that of neutron stars in scalar-tensor
theories~\cite{Damour:1996ke}.
In the Gaussian model, it would be particularly interesting to compute the
sensitivities of the spin-induced scalarized black holes of
Refs.~\cite{Herdeiro:2020wei,Berti:2020kgk}.

\appendix

\section*{Acknowledgments}
%
We thank Carlos~A.~R.~Herdeiro, Mohammed~Khalil, Eugen~Radu,
Jan~Steinhoff, and Helvi~Witek for numerous discussions.
We also thank Alessandra Buonanno and Harald Pfeiffer for questions that helped
us improve parts of the text.
H.O.S and N.Y. are supported by NASA Grants No.~NNX16AB98G and No.~80NSSC17M0041.
N.Y. also acknowledges support from the Simons Foundation through Award number 896696.
F.-L.J. and E.B. are supported by NSF Grants No. PHY-1912550,
AST-2006538, PHY-090003 and PHY-20043, and NASA Grants
No. 17-ATP17-0225, 19-ATP19-0051 and 20-LPS20-0011.
The figures in this work were produced with {\sc Matplotlib}~\cite{Hunter:2007}
and {\sc TikZ}~\cite{tantau:2021}.
This work has received funding from the European Union's Horizon 2020 research
and innovation programme under the Marie Skłodowska-Curie grant agreement No.~690904
and networking support by the GWverse COST Action CA16104,
``Black holes, gravitational waves and fundamental physics''.

\begin{figure*}[ht]
\includegraphics[width=\columnwidth]{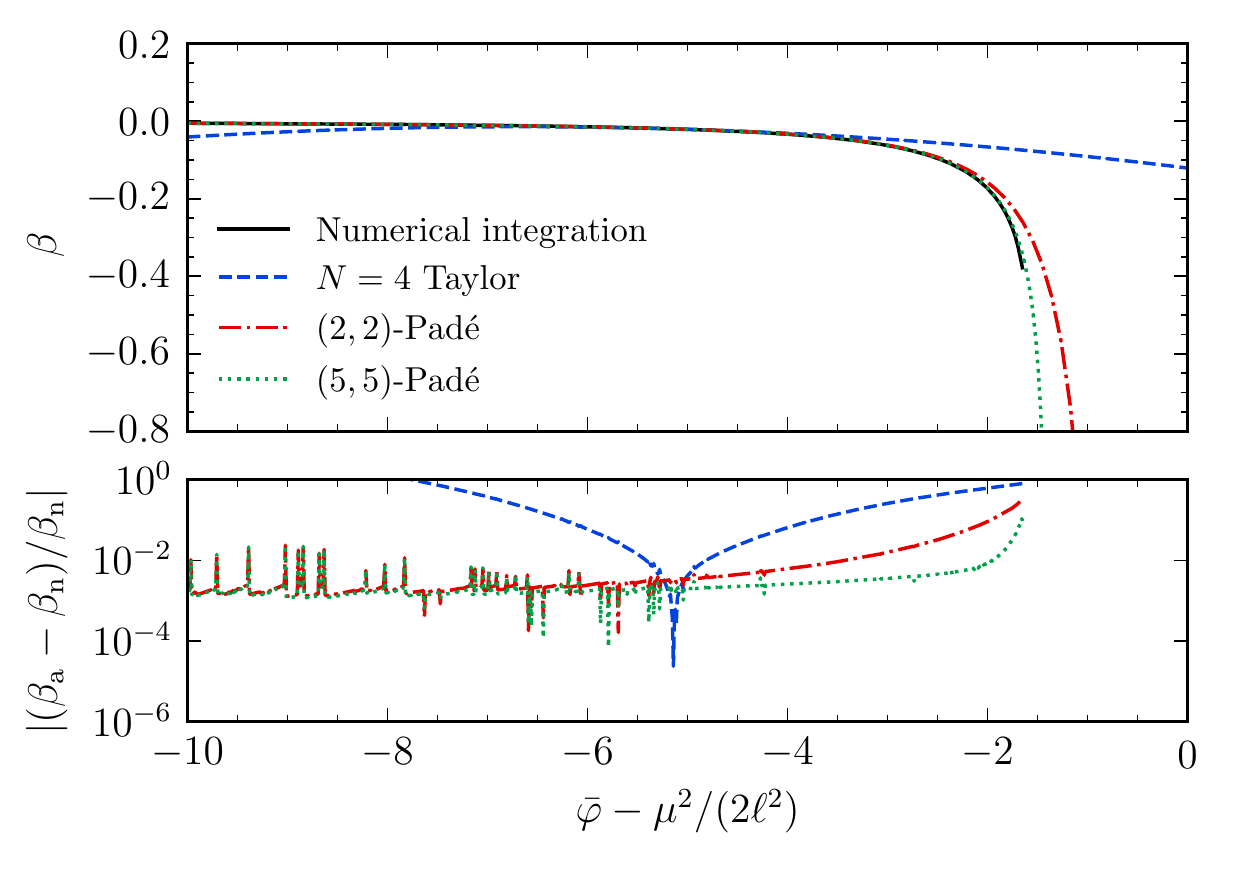}
\includegraphics[width=\columnwidth]{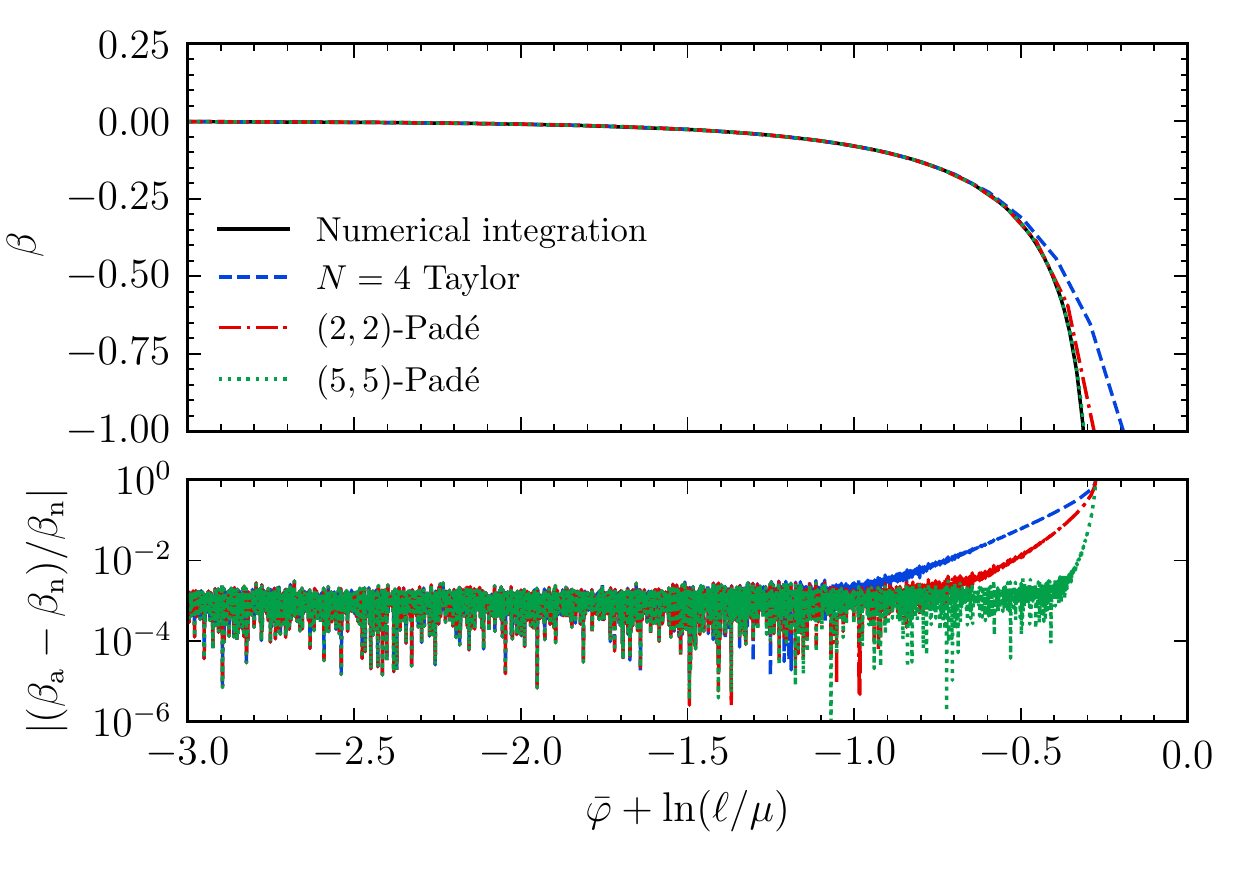}
\caption{
Black hole sensitivities $\beta$ obtained from $\alpha$ in
Figs.~\ref{fig:ss_sensitivity_vs_phi0} and~\ref{fig:dil_sensitivity_vs_phi0}
using the definition~\eqref{eq:sensitivity_beta}.
Left panel: the sensitivity $\beta$ in the shift-symmetric case as a function
of $\vpbg - \mu^{2} / (2\lsqr)$, cf. Eq.~\eqref{eq:combinationShiftSym}.
Right panel: the sensitivity $\beta$ in the dilatonic case as a function of
$\vpbg + \ln(\ell / \mu)$ cf. Eq.~\eqref{eq:combinationDilaton}.
The upper panels show numerical and analytic results obtained from
Eq.~\eqref{eq:analytical_sens} with $N=4$, its $(2,2)$-Pad\'e resummation, and
the $(5,5)$-Pad\'e resummation of Eq.~\eqref{eq:analytical_sens} with $N=10$.
The bottom panels show the fractional error between analytic (``a'') and
numerical (``n'') calculations.
The numerical sensitivities and their ($5,5$)-Pad\'e counterpart show excellent
agreement, except for one qualitative difference: the Pad\'e approximants are
singular, while the numerical curves end at $\vpbg - \mu^{2} / (2\lsqr) \approx
-1.651$ and $\vpbg + \ln(\ell / \mu)\approx -0.276$ as one approaches the saturation of
the theories' respective horizon bounds~\eqref{eq:ss_max_scalar} and~\eqref{eq:dl_bound}.
In the limit $\vpbg\to -\infty$ we have $\beta\to 0$ for both theories, and at
the end points we find $\beta=-0.376$ and $\beta=-1446$, respectively.
}
\label{fig:sensitivity_beta_ss_and_dil}
\end{figure*}

\begin{figure*}[ht]
\includegraphics[width=\columnwidth]{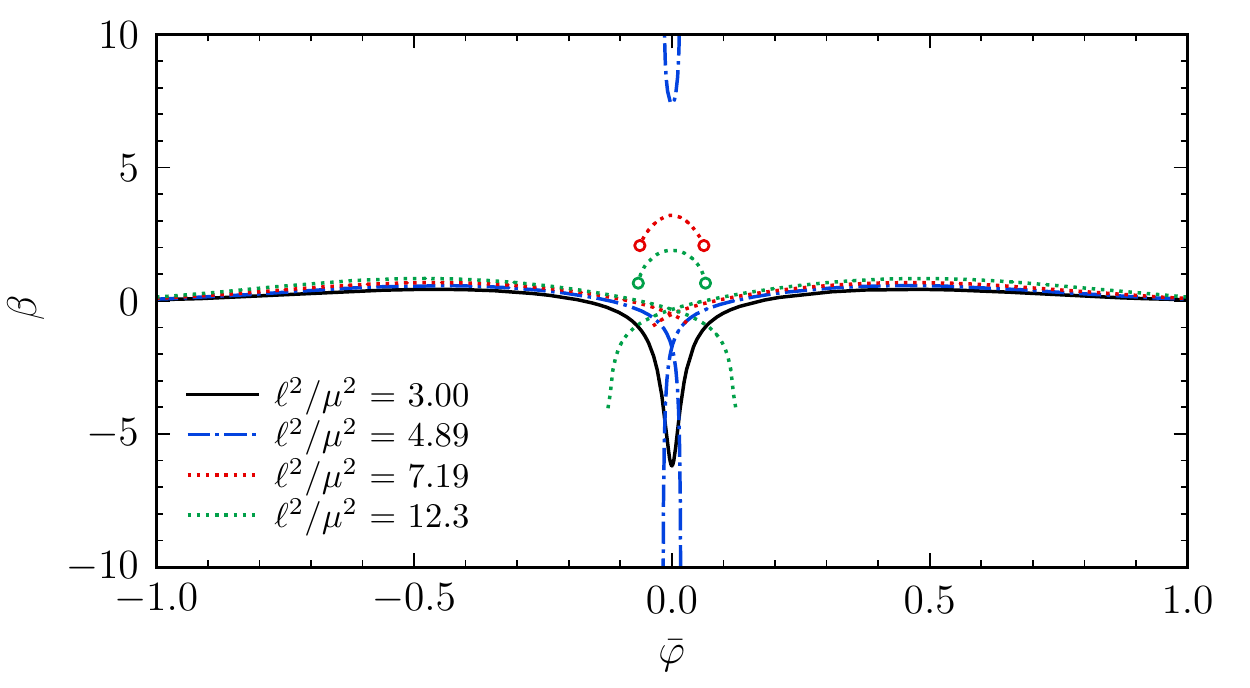}
\includegraphics[width=\columnwidth]{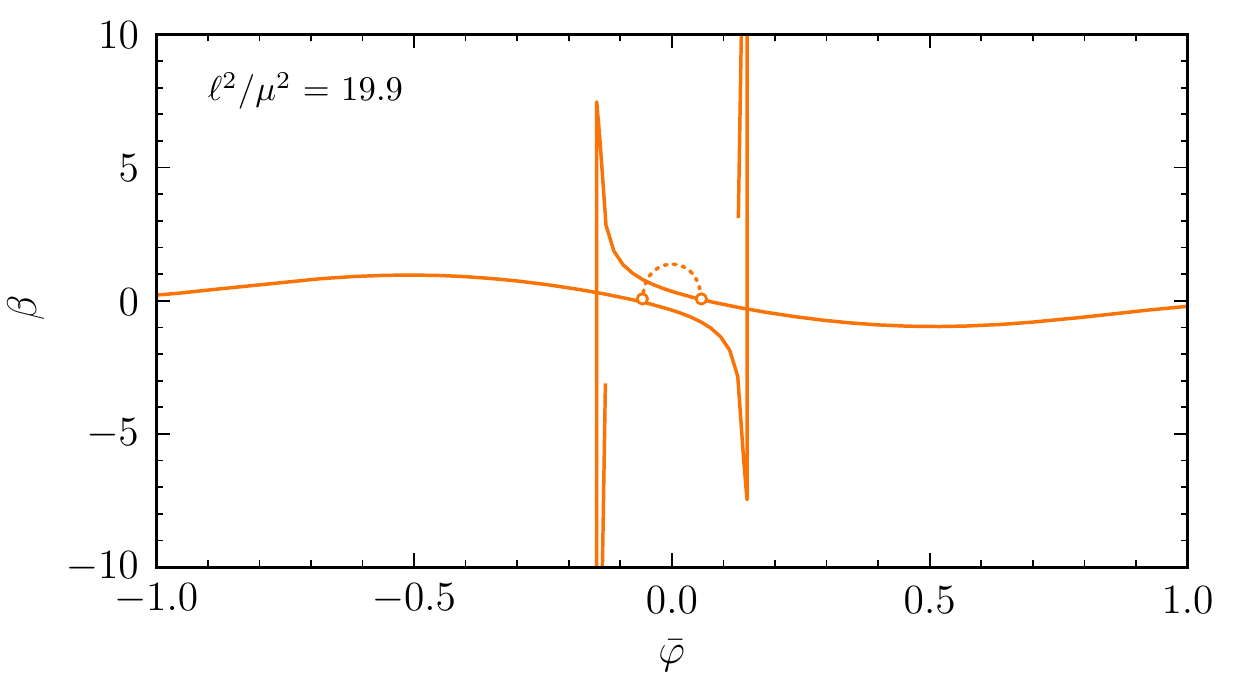}
\caption{
Black hole sensitivities $\beta$ in the Gaussian theory as
functions of $\vpbg$ and $\ell/\mu$. We consider the same constant entropy
solutions as in Fig.~\ref{fig:gaussian_l6_sens_trans}.
For values $(\ell/\mu)^2\lesssim 3.82$ the curve includes a stable Schwarzschild solution at $\vpbg=0$, with $\alpha=0$, but $\beta< 0$ can be large, cf. $(\ell/\mu)^2=3.00$.
When $(\ell/\mu)^2 \gtrsim 3.82$ the curve is multivalued.
The segments bracketed by the circles denote the ``Schwarzschild branches'' with $\beta> 0$. They
include an unstable Schwarzschild solution at $\vpbg=0$, with $\alpha=0$.
The remainder of the curves form two ``scalarized branches'' that include the
stable scalarized black holes with $\vpbg = 0$ and $\alpha\neq 0$.
For $(\ell/\mu)^2\in[3.82, 5.61]$, the three branches are connected.
At their junction, $M/\mu$ features a cusp, cf. Fig.~\ref{fig:gaussian_l6_sens_trans}, and the slope of $\alpha$, that is $\beta$ here, is infinite.
When $(\ell/\mu)^2\gtrsim 5.61$, the branches are discontinuous due to the violation of the horizon bound~\eqref{eq:reg_gauss_rH}.
But when $(\ell/\mu)^2\in[13.6,13.8]$ and $(\ell/\mu)^2\gtrsim 14.0$, the ratio $M/\mu$ of scalarized branches also features a cusp and $\beta$ hence diverges. This
is illustrated with~$(\ell/\mu)^2=19.9$ in the right panel.
}
\label{fig:sensitivity_beta_gauss}
\end{figure*}

\begin{figure}[t]
\includegraphics[width=\columnwidth]{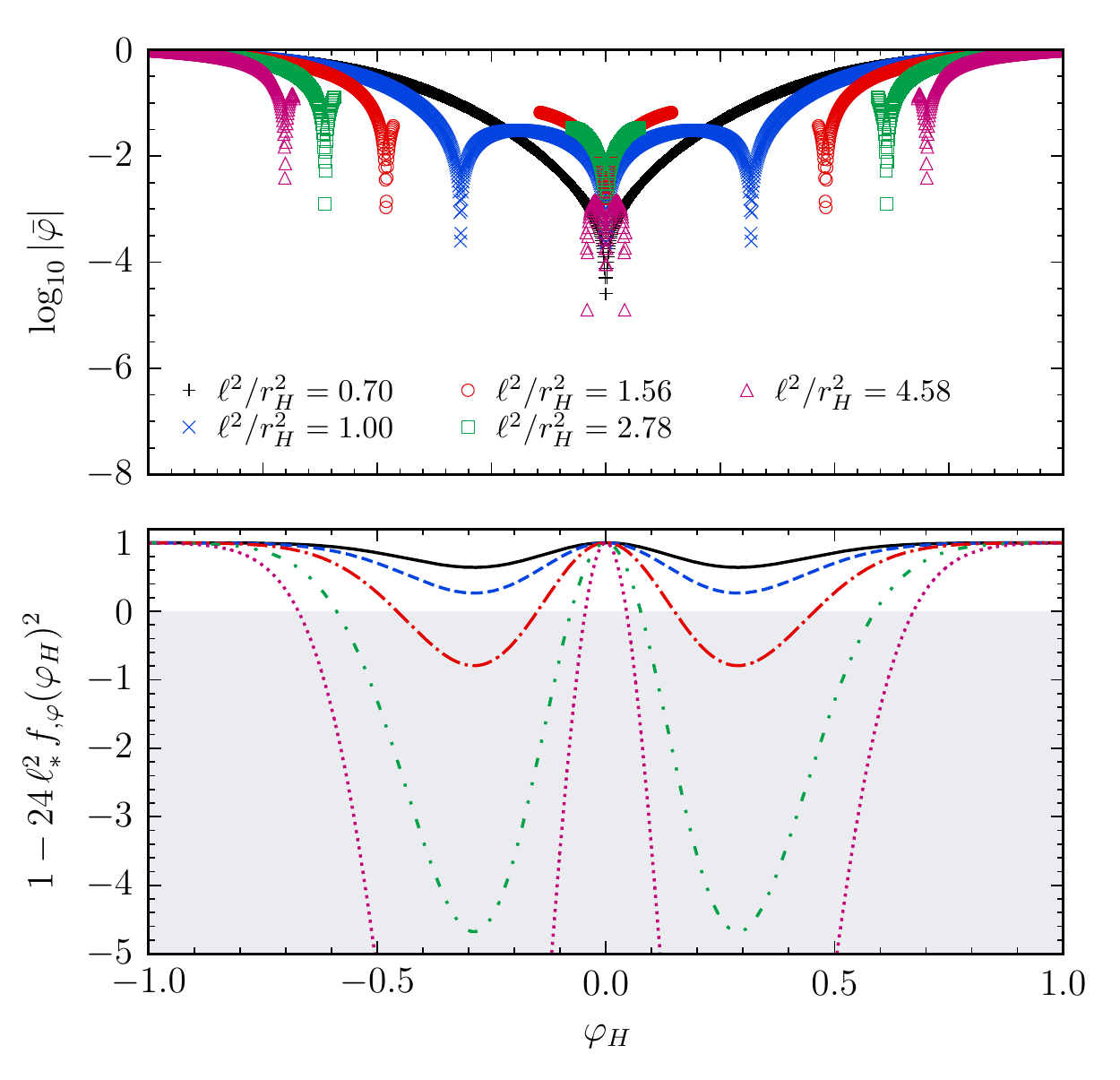}
\caption{
Finding scalarized black hole solutions in the Gaussian theory~\eqref{eq:coupling_gauss}.
Top panel: the asymptotic value $\log_{10}|\vpbg|$ of the scalar field as a function of its value
at the horizon $\vp_H$ for several ratios $\lstar=\ell/r_H$.
When $\lstar^2 \lesssim 0.725$ only one zero of $\vpbg$, located at $\vp_H = 0$ exists as shown by the cusps in the
data points. It corresponds to the Schwarzschild solution.
As we increase $\lstar$, two additional zeros of $\vpbg$ with $\vp_H \neq 0$ appear.
They have the same magnitude, but opposite signs, as expected from the theory's $\mathds{Z}_{2}$-symmetry.
Note that for $\lstar^2 = 1.56$, $2.78$ and $4.58$ the curve is not continuous.
Bottom panel: the existence condition~\eqref{eq:scalarh_bound_no_sw} as a function of $\vp_H$.
The solid, dashed, dot-dashed, dash-double-dotted and dotted lines correspond respectively to
$\lstar^2 = \{0.7,\, 1.0,\, 1.56,\, 2.78,\, 4.58\}$.
In the shaded region, Eq.~\eqref{eq:scalarh_bound_no_sw} is violated.
For $\lstar^2 = 1.56$, $2.78$ and $4.58$, the condition is violated on $\vp_H$ intervals.
This causes the discontinuity in the data points with the same values of $\lstar$ in the top panel.
}
\label{fig:find_scal_sols}
\end{figure}

\section{Near-horizon expansion of the Ricci and Kretschmann curvature invariants}
\label{app:RicciKretschmann}

In Section~\ref{subsec:nonrotating_bhs}, we obtained the coefficients of the power series expansions~\eqref{eq:horizon_exp} up to $\varphi_4^H$, $N_4^H$ and $\sigma_3^H$. We then computed the scalar field and Gauss-Bonnet invariant as in Eqs.~\eqref{nearHorizonPhiAndGB}. We can use the same coefficients to calculate the Ricci and Kretschmann curvature invariants $R$ and $\mathcal{K}=R^{\mu\nu\rho\sigma}R_{\mu\nu\rho\sigma}$ as:
\begin{subequations}
\begin{align}
R \, r_H^2 &=\rho_{H}+ \sum_{n=1}^{2}\rho^{H}_n (r_{\ast} - 1)^n + \mathcal O(r_\ast-1)^3\,,\label{nearHorizonR} \\
\mathcal{K} \, r_H^4&=k_H+\sum_{n=1}^{2} k^H_n (r_{\ast} - 1)^n + \mathcal O(r_\ast-1)^3\,,\label{nearHorizonK}
\end{align}\label{nearHorizonRandK}%
\end{subequations}
where
the coefficients are long functions of $\lstar$ and $\varphi_H$ available online~\cite{FLJRepo}.
Near the saturation of the horizon bound \eqref{eq:scalarh_bound_no_sw}, i.e. for $\epsilon^2=1-24\,\lstar^{4}\, f_{,\vp}(\vp_H)^2\ll 1$, we find
\begin{subequations}
\begin{align}
\rho_H &= 2+\mathcal O(\epsilon)\,,\label{eq:RicciHorizonCoeffs1} \\
\rho^{H}_1 &= -18\frac{\chi}{\epsilon}+\mathcal O(\epsilon^{0})\,, \\
\rho^{H}_2 &= \frac{243}{16}\frac{\chi^2}{\epsilon^3}+\mathcal O(\epsilon^{-2})\,,
\end{align}\label{eq:RicciHorizonCoeffs}%
\end{subequations}
and
\begin{subequations}
\begin{align}
k_H&=84+\mathcal O(\epsilon)\,,\label{eq:KretschHorizonCoeffs1}\\
k^{H}_1&=-648 \frac{\chi}{\epsilon}+\mathcal O(\epsilon^{0})\,,\\
k^{H}_2&=\frac{2187}{4}\frac{\chi^2}{\epsilon^3}+\mathcal O(\epsilon^{-2})\,,
\end{align}\label{eq:KretschHorizonCoeffs}%
\end{subequations}
with $\chi=3+4 \lstar^{2} f_{,\vp\vp}(\vp_H)$. As with the Gauss-Bonnet scalar, we have that $\rho_H$ and $k_H$ are finite and do not depend on $f(\varphi)$ in this limit,
while the other coefficients in Eqs.~\eqref{eq:RicciHorizonCoeffs}-\eqref{eq:KretschHorizonCoeffs} are singular. The near-horizon expansion of the curvature invariant $R^{\mu\nu}R_{\mu\nu}=(\mathcal{K}+R^2-\sG)/4$ can then be inferred from our results, and its first term is finite too.

\section{Numerical methods}
\label{app:numerical_methods}

For all our numerical calculations, we used {\sc Mathematica}'s differential equation solving function
{\sc NDSolve}, with the method ``{\sc StiffnessSwitching}'', that automatically changes between
a nonstiff or stiff solver when necessary.
We set both {\sc PrecisionGoal} and {\sc AccuracyGoal} to 15, and worked with
the default {\sc WorkingPrecision}.
The integrations of Eqs.~\eqref{eq:eom_solve_num} were performed in the domain $r_{\ast} \in [1 - 10^{-\epsilon},\, 10^{10}]$, with $\epsilon = 5$.
An exception is in the near-horizon integrations done in Sec.~\ref{sec:coup_ssym} cf. Fig.~\ref{fig:ss_sensitivity_vs_phi0_bh_props}.
There we set {\sc WorkingPrecision} to machine precision and $\epsilon = 6$.

To calculate the asymptotic parameters $M_{\ast}$, $Q_{\ast}$ and $\vpbg$ in Eqs.~\eqref{eq:asympt_expansions}, we proceeded as follows.
First, from the numerical integration we know the values of $\vp$, $\vp'$, and $N$
at our ``numerical infinity'', $r_{\ast} = 10^{10}$.
Then, the value $\vp$ at $r = r_* = 10^{10}$ gives $\vpbg$, since for $r_{\ast} \gg 1$ all $1/r_{\ast}$ corrections are negligible.
Next, the values of $N$ and $\vp'$ are respectively used in the right-hand sides of Eq.~\eqref{eq:asympt_expansion_n}
and of Eq.~\eqref{eq:asympt_expansion_vp} (after taking a derivative with respect to $r_{\ast}$).
This constitutes a system of two equations for the two unknowns $M_{\ast}$ and $Q_{\ast}$, which is then
solved with {\sc Mathematica}'s {\sc NSolve} function.
As a consistency check, we verified that $M_{\ast}$ calculated this way agrees with the directly evaluation of Eq.~\eqref{eq:def_n} at $r_{\ast} = 10^{10}$.

\section{Black hole sensitivity $\beta$}
\label{app:beta_sensitivity}

We gather here the sensitivities $\beta$ of black holes in the shift-symmetric,
dilatonic, and Gaussian theories, obtained from the numerical and analytic
sensitivities $\alpha$ of
Figs.~\ref{fig:ss_sensitivity_vs_phi0},~\ref{fig:dil_sensitivity_vs_phi0}
and~\ref{fig:gaussian_l6_sens_trans} using Eq.~\eqref{eq:sensitivity_beta}
(recall that a fixed $\sS_{\rm W}$ is equivalent to a fixed $\mu$).
They are useful in the context of PN calculations. For instance, they enter the
1PN Lagrangian of Ref.~\cite{Julie:2019sab}.

In Fig.~\ref{fig:sensitivity_beta_ss_and_dil} we show $\beta$
in the shift-symmetric (left panel) and dilatonic (right panel) cases.
We see once more the remarkable agreement between the numerical sensitivities
and their $(5,5)$-Pad\'e counterparts.
For a black hole with fixed irreducible mass $\mu$ in the shift symmetric
case, we find $\beta\to0$ for $\vpbg\to-\infty$ and $\beta=-0.376$ at the
end point.
In the dilatonic case, we have $\beta\to0$ for $\vpbg\to-\infty$ and $\beta
\approx -1446$ at the end point.

In Fig.~\ref{fig:sensitivity_beta_gauss}, we show $\beta$ in the Gaussian case,
for the $\ell/\mu$ values chosen in Fig.~\ref{fig:gaussian_l6_sens_trans}.
When $(\ell/\mu)^2 \lesssim 3.82$, the spontaneous scalarization of
Schwarzschild black holes ($\vp=0$) does not occur, and we found $\alpha=0$ at
$\vpbg=0$ in Fig.~\ref{fig:gaussian_l6_sens_trans}.
By contrast, the sensitivity $\beta$ of these stable Schwarzschild black holes
is nonzero at $\vpbg=0$, and it can even be large and finite,
cf.~$(\ell/\mu)^2=3.00$ in Fig.~\ref{fig:sensitivity_beta_gauss}.
Above the scalarization threshold, a Schwarzschild branch with $\beta>0$
bracketed by the circles appears, together with two scalarized branches.
Given the definition \eqref{eq:sensitivity_beta}, the sensitivities $\beta$ of
the latter are even-symmetrical due to the theory's $\mathds{Z}_{2}$ symmetry.
We recall that when $3.82\lesssim(\ell/\mu)^2\lesssim 5.61$, the branches are
connected. As shown, e.g., for the example $(\ell/\mu)^2=4.89$ in
Fig.~\ref{fig:gaussian_l6_sens_trans}, $M/\mu$ features a cusp at their
junction.
This means that the slope of $\alpha$, i.e. $\beta$ in
Fig.~\ref{fig:sensitivity_beta_gauss}, is infinite there.
When $(\ell/\mu)^2\gtrsim 5.61$, the branches are discontinuous, cf. $(\ell /
\mu)^2 = 7.19$ and $(\ell / \mu)^2 =12.30$ in
Fig.~\ref{fig:sensitivity_beta_gauss}.
The discontinuity happens due to the existence of $\vp_H$ ranges that
do not satisfy the horizon bound~\eqref{eq:reg_gauss_rH}.
Our results for values $(\ell/\mu)^2\leqslant 20$ in $\Delta(\ell/\mu)^2 \approx 0.2$
increment can be found in~\cite{FLJRepo}.

\section{Obtaining spontaneously scalarized black holes}
\label{app:find_scalarized}

We briefly review here how spontaneously scalarized black hole solutions have
been obtained in the literature \cite{Doneva:2017bvd,Silva:2017uqg} for the
example of the Gaussian theory~\eqref{eq:coupling_gauss}, when the scalar field
vanishes asymptotically.

We first choose a pair of values $\lstar=\ell/r_H$ and $\vp_H$, and
numerically integrate Eqs.~\eqref{eq:eom_solve_num} outwards, from $r_{\ast} = r/r_H = 1$ up to
a large value of $r_{\ast}$, and
we extract the asymptotic scalar field value $\vpbg$.
We repeat these steps for a range of $\vp_H$ values
allowed by the reality condition~\eqref{eq:scalarh_bound_no_sw},
while keeping $\lstar$ fixed.
The outcome is a function $\vpbg (\vp_H)$ that
generally has a single zero at $\vp_H = 0$ corresponding to the
Schwarzschild solution [cf. below Eq.~\eqref{eq:coupling_gauss}].
However, for certain disjoint $\lstar$ ranges, an additional even number of
zeros with equal and opposite $\vp_H \neq 0$ appear.
They correspond to scalarized black holes, which come in pairs due to
the theory's $\mathds{Z}_{2}$-symmetry.

The pair of solutions with smallest $|\vp_H|_0$
values has a nodeless scalar field configuration
(``ground state''), while solutions with successively increasing $|\vp_H|_{k}$ values correspond to scalar field configurations with $k$
nodes (``excited states'').
In the first $\lstar$ range (with smallest $\lstar$ values) allowing for spontaneous scalarization, only ground states with $k=0$ are found.
In the second $\lstar$ range, $k=0$ states and their excited $k=1$ counterparts are observed.
In the third $\lstar$ range, $k=0,1,2$ states are observed, and so on.
It must however be noted that excited states with $k\geqslant1$ are radially unstable~\cite{Blazquez-Salcedo:2018jnn}.

In the present paper and in the online repository~\cite{FLJRepo} we focus on $\ell/\mu$ ratios up to $(\ell/\mu)^2= 20$, for which scalarized black holes with $\vpbg=0$ have $\vp_H\approx\pm 0.70$.
This translates into $\lstar^2\lesssim 4.59$ using Eq.~\eqref{eq:a_mu_ratio}.
For such $\lstar$ values, Ref.~\cite{Blazquez-Salcedo:2018jnn,Blazquez-Salcedo:2020rhf} showed that only ground states exist, hence the presence of at most one pair of
nonzero sensitivities $\alpha$ at $\vpbg=0$ in Fig.~\ref{fig:gaussian_l6_sens_trans}.
Moreover Refs.~\cite{Blazquez-Salcedo:2018jnn,Blazquez-Salcedo:2020rhf} proved that ground states are always radially and axially
stable, when $\vpbg=0$ if $\lstar^2\lesssim 25.02$.
This implies that our scalarized black holes are stable at least when $\vpbg=0$.

Fig.~\ref{fig:find_scal_sols} shows $\log_{10}|\vpbg|$ as a function of
$\vp_H$ (top
panel) and the regularity condition~\eqref{eq:scalarh_bound_no_sw} (bottom panel) for $\lstar^2 = \{0.7,\, 1.0,\, 1.56,\, 2.78,\, 4.58\}$.
The smallest $\lstar$ values are respectively slightly below the
scalarization threshold $\lstar^2 \approx 0.725$, while the other
values are those of Table~\ref{tab:scalarized_bhs_props}.

\bibliography{biblio}

\end{document}